\documentclass{aa}
\usepackage{graphicx}
\usepackage{natbib}
\usepackage{txfonts}
\usepackage{rotating}
\begin{document}

\def\kmsmpc{km~s$^{-1}$~Mpc$^{-1}$}
\def\ho{H$_0$}
\newcommand{\degree}{\ensuremath{^\circ}}

\title{On the evolution of environmental and mass properties of strong lens galaxies in COSMOS \thanks{Based on observations made with ESO telescopes
    at Paranal observatory under program ID 077.A-0473(A) and under
    large program ID 175.A-083911.}}

 \subtitle{}

 \author{C. Faure\inst{1} \and T. Anguita\inst{2,3}  \and D. Alloin\inst{4} \and K. Bundy \inst{5} \and  A. Finoguenov\inst{6} \and A. Leauthaud
   \inst{7,8}   \and   C. Knobel\inst{9} \and J.-P. Kneib\inst{10} \and E. Jullo\inst{11} \and O. Ilbert\inst{10}
   \and A. M. Koekemoer\inst{12}  \and  P. Capak\inst{13} \and N. Scoville\inst{13} \and L.A.M. Tasca\inst{10} }

\institute{
Laboratoire d'Astrophysique, Ecole Polytechnique F\'ed\'erale de Lausanne (EPFL), Observatoire de Sauverny, CH-1290 Versoix, Switzerland
\and 
Centro de Astro-Ingenier\'ia, Departamento de Astronom\'ia y Astrof\'isica, Pontificia Universidad Cat\'olica de Chile, Casilla 306, Santiago, Chile
\and
Max Planck Institut f\"{u}r Astronomia, K\"{o}nigstuhl 17, D-69117, Heidelberg, Germany
\and
Laboratoire AIM, CEA/DSM-CNRS-Universite Paris Diderot, IRFU/SEDI-SAP, Service d'Astrophysique, CEA Saclay, Orme des Merisiers, 91191 Gif-sur-Yvette, France
\and
Hubble Fellow, Department of Astronomy, UC Berkeley, 601 Campbell Hall, Berkeley, CA 94720-3411, USA
\and
Max-Planck Institut f\"{u}r Extraterrestrische Physik, Giessenbachstrasse, 85748, Garching, Germany 
 \and
 Lawrence Berkeley National Laboratory, 1 Cyclotron Road, Berkeley CA 94720
\and
Berkeley Center for Cosmological Physics, University of California, Berkeley, CA 94720, USA
\and
Institute for Astronomy, ETH Zurich, Wolfgang-Pauli-Strasse 27, 8093 Zurich, Switzerland
\and
Laboratoire d'Astrophysique de Marseille, CNRS Universit\'e de Provence, 38 rue F. Joliot-Curie, 13388 Marseille Cedex 13, France     
\and
Jet Propulsion Laboratory, MS 169-327,4800 Oak Grove Drive, Pasadena, CA 91109, USA
\and
Space Telescope Science Institute, 3700 San Martin Drive, Baltimore, MD 21218, USA
\and
California Institute of Technology, MC 105-24, 1200 East California Boulevard, Pasadena, CA 91125, USA}
\date{Received 20 October 2009 / Accepted 09 February 2011 }
\authorrunning{C. Faure et al.}
\titlerunning{Environment and  mass properties of COSMOS lens galaxies.}

\abstract{Nearly 100 new strong lens candidates have been discovered
  in the COSMOS field. Among these, 20 lens candidates with
  0.34$\lesssim z_{lens} \lesssim $1.13, feature multiple images of
  background sources.}  {Using the multi-wavelength coverage of
  the field and its spectroscopic follow-up, we
  characterize the evolution with redshift of the  environment  and of the dark-matter (DM) fraction of the lens galaxies.}{We present spectroscopic and new photometric
  redshifts of the  strong lens candidates. The lens environment is
  characterized in the following way: we account for the projected
  10 closest galaxies around each lens and for galaxies with a
  projected distance less than 1~Mpc at the lens galaxy redshift. In both cases,  we perform similar measurements on a control
  sample of "twin" non-lens early type galaxies (ETGs). In addition, we identify group members and field galaxies in the X-ray and optical catalogs of galaxy groups and clusters.  From those catalogs, we measure the external shear contribution of the groups/clusters surrounding the lens galaxies. The systems are then modeled using  a SIE for the lens galaxies plus the external shear due to the groups/clusters. }{We observe that the average stellar
  mass of lens galaxies increases with redshift. In addition, we measure that the environment of lens galaxies is compatible with that of the
  twins over the whole redshift range tested here. During the lens modeling, we notice that, when let free, the external shear points in a direction which is the mean direction of the external shear due to  groups/clusters and of  the closest galaxy to the lens.
 We also notice that the DM fraction of the lens galaxies measured within the Einstein radius significantly decreases as the redshift increases.  }{Given these, we conclude that, while the environment of lens galaxies is compatible with that of non-lens ETGS over a wide range of redshifts, 
  their mass properties evolves significantly with redshift: it is still
  not clear whether this advocates in favor of a stronger lensing bias toward massive objects at high redshift or is simply representative of
  the high proportion of massive and high stellar density galaxies  at high redshift.
   }

\keywords{Cosmology: Gravitational lensing}
\maketitle

\section{Introduction}
In the field of gravitational lensing, studies of strong galaxy-galaxy
lenses have recently become possible on statistical grounds. Indeed, within just a few years,
searches in the SDSS (SLACS: Bolton et al. 2006, 2008, Allam et
al. 2007), in COSMOS (Faure et al. 2008, hereafter Paper~I; Jackson 2008) and in the
CFHT-LS surveys (SL2S: Cabanac et al. 2007, Limousin et al. 2009b)
have delivered more than two hundred new strong galaxy-galaxy lenses,
spanning a wide range in redshift and image angular separation. The
reasons for this interest are several. First, strong lens
galaxies provide measurements of the total mass distribution on galaxy
scales, bringing additional information about the processes of galaxy
formation and evolution (e.g. Ofek et al. 2003, Chae et al. 2006,
Koopmans et al. 2006, Tortora et al. 2010). Second, from a statistical point of view,
strong lensing occurrences trace the abundance and concentration of
matter in the universe, providing another test for cosmological models
(e.g. Keeton 2001, Bartelmann et al. 2003).

Studies of the physical and environmental properties of low redshift
strong lens galaxies (z$\lesssim$0.3) have shown that they are {\it
  bona-fide} massive ETGs (Treu et al. 2006, 2009 -T09 hereafter-, Gavazzi et al. 2007, Grillo et al. 2009). The strong lens sample in
the COSMOS field (Paper~I), spans a
higher redshift range (0.33$\leq$z$\leq$1.13) and previous
work on this sample have shown that the projected distribution of lens galaxies in
COSMOS is comparable to that of ETGs, whether in rich
environments such as cosmic filaments, or in the field
(Faure et al. 2009, hereafter Paper~II). Interestingly, this result  means that multiple-mass-sheets do not contribute significantly to make more efficient lenses, contrary to results found in ray tracing through numerical simulations (Wambsganss et al. 2005, Hilbert et al. 2007, 2008). In Paper~II we also demonstrate that the
presence of large scale structures (LSS) in the lens galaxy environment 
increases the angular separation of the lensed images of a source, as
noticed earlier by Oguri et al. (2005).

 In addition, analyses of the mass density profiles of strong lens galaxies from
hydrodynamical N-body simulations (Dobke et al. 2007, Limousin et al.
2009a) and lens samples (Limousin et al. 2007, Auger 2008, Natarajan et al. 2009) show that the mass distribution in
galaxies depends on their environment. For example, compared to
galaxies centrally placed in their group or cluster, some skewing of
the total mass density slope appears in galaxies located at the edges
of a group or cluster. This is consistent with tidal stripping of
their dark matter halo (T09).

 Therefore, it is important to explore the local environment of the COSMOS strong lenses rather than LSS as in Paper~II. We  also want to learn about the lens galaxy mass  properties, in particular to derive information relative to their DM fraction (f$_{DM}$, defined as f$_{DM}$=1-M$_\star$/(M$_\star$+M$_{DM}$). This is possible as the COSMOS strong lenses pertain to  one of
the best studied region of the sky, for which deep multi-wavelength
observations have been collected (Koekemoer et al. 2007, Taniguchi et
al. 2007, Capak et al. 2007a, Scoville et al. 2007, Lilly et
al. 2007, Ilbert et al. 2009). Thus the properties of the COSMOS galaxy population have
been studied extensively (e.g. McCracken et al. 2007, Capak et
al. 2007b, Scarlata et al. 2007a, 2007b, Ilbert et
al. 2010). 

The paper is organized as follows. In $\S$~\ref{syst}, we provide improved redshifts and stellar masses of the COSMOS lens galaxies. In
$\S$~\ref{densityratios}, we examine and discuss the lens
environment. In $\S$~\ref{model}, we present results from our
strong lens mass models for the sub-sample of
triple and quadruple image systems. Discussion and conclusions appear in
$\S$~\ref{disc}. Throughout this paper, we assume a WMAP5 $\Lambda$CDM
cosmology with $\Omega_{\rm m}=0.258$, $\Omega_\Lambda=0.742$, $H_0=72$ $h_{72}$~km~s$^{-1}$~Mpc$^{-1}$.

\section{The COSMOS sample of strong lenses: new redshifts and stellar masses}\label{syst}

By visual inspection of stamp images of 10\arcsec$\times$10\arcsec\,
around $\sim$9500 early photometric type galaxies (with redshifts : 0.2~$\leq$~z~$\leq$~1.0, and absolute magnitude: M~$_{\rm V}<$~-20) in the COSMOS field,
we have discovered 60 strong lens candidates (Paper I). In addition, in the same study, 7 lens candidates were found serendipitously across the field. Among the whole sample,
19 systems display long curved arcs or multiple images with similar
colors (based on SUBARU images, by Taniguchi et al. 2007) around a bright
lens galaxy, and with an image arrangement around the lens galaxy
which is consistent with the lensing hypothesis, as probed by lens
modeling.

Independently, Jackson (2008) inspected the complete set of
galaxies in the COSMOS field, without discrimination. He has
discovered two additional convincing strong lens candidates producing
multiple images: J095930.93+023427.7 and J100140.12+020040.9. The lens galaxies both appear to be ETGs.

Among the 21 multiple image lenses now available, one system
(COSMOS~5921+0638) -- being the only confirmed lensed quasar of the
sample -- has been studied separately by Anguita et al. (2009,
hereafter Paper~III). Another system, COSMOS~5737+3424, is a
galaxy cluster that is lensing a set of background galaxies. In such a
case, the lens covers a different mass scale than in the other 20
systems, and the lens potential is more complex than a single lens
galaxy: this target has been dropped from the present analysis.

\subsection{Spectroscopic followup}\label{obs}

\begin{figure}[ht]
\begin{center}
\includegraphics[width=8cm]{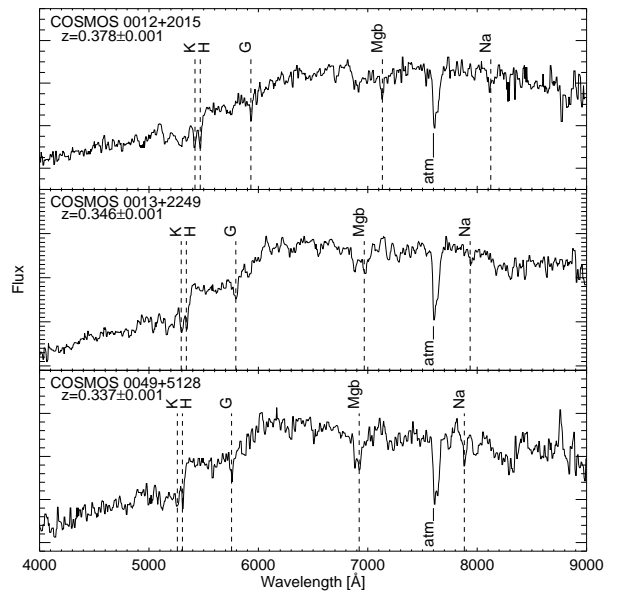}
\caption{VLT/FORS1 spectra of three strong lens candidates. The spectral resolution is 5.52~\AA\, per pixel, and has been smoothed by a 3 pixels box to improve the quality of the display. Flux is in arbitrary units.}
\label{FORS_spec}
\end{center}
\end{figure}
\begin{figure}[ht]
\begin{center}
\includegraphics[width=8cm]{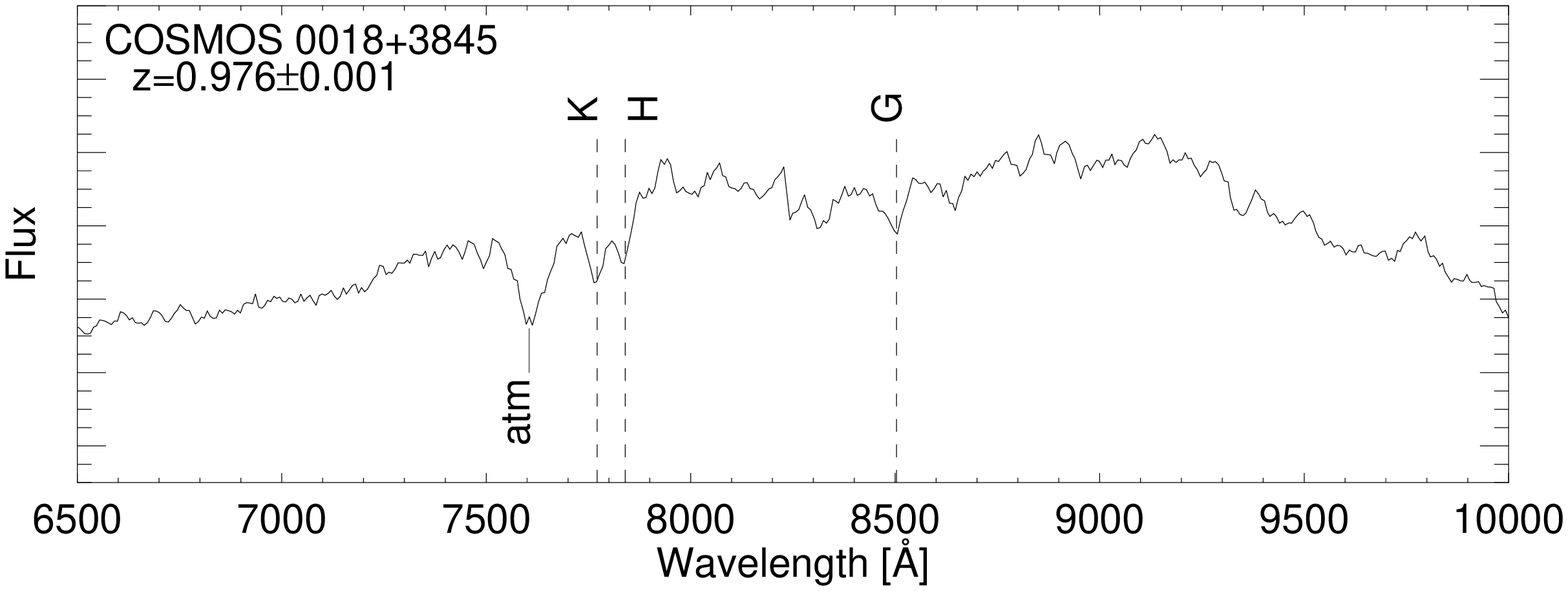}
\caption{Keck/DEIMOS spectrum of COSMOS~0018+3845. The spectrum is smoothed by 2.3\AA\ and binned down to 7.5\AA\ per pixel, taking the wavelength dependent noise into account, to improve the apparent signal-to-noise. Flux is in arbitrary units.}
\label{deimos_spec}
\end{center}
\end{figure}
On March 3$^{\rm rd}$ and April 25-26$^{\rm th}$ 2006, we have
successfully obtained spectra for five of the strong lenses in the COSMOS
field (PI: Faure) using the FORS1 instrument (ESO/VLT) in multi-object
spectroscopic mode with 19 slits. For each target, the central
slit was located on the lens candidate with an orientation
intersecting both the lens galaxy and the brightest lensed image. The
other 18 slits in the $\sim$7'x7' field around the lens, were
preferentially positioned on galaxies with colors similar to that of
the lens galaxy (likely at the same redshift). Leaving aside
COSMOS~5921+0638 (discussed in Paper~III) and the lens cluster COSMOS~5737+3424, we display
the spectra of the other three lens galaxies in
Fig.~\ref{FORS_spec}. Regarding
the source redshifts, from this dataset we could extract only one
source redshift, in COSMOS~5921+0638 (Paper~III). Positions and
redshifts of  secondary targets around COSMOS~0012+2015, COSMOS~0013+2249 and COSMOS~0049+5128 are given in Table~\ref{tabexan}. 

 In addition, several COSMOS lens candidates have been observed with the Keck telescopes.
 Using the Deep Extragalactic Imaging Multi-Object Spectrograph (DEIMOS) a spectrum of COSMOS~0018+3845 has been obtained on February 11, 2010 (Fig.~\ref{deimos_spec}). The data were collected in 7$\times$1800s exposures under photometric conditions with 0.5-0.7\arcsec\, seeing.  We used a 1\arcsec\, slit together with  the 830 line/mm grating tilted to 7860\AA\, and the OG550 blocking filter. The resulting spectral resolution is $<3.3$\AA, depending on seeing and object morphology. The objects were dithered along the slit by $\pm3$\arcsec\, between exposures to improve background subtraction, as described in Capak et al. (2009).  The redshift of the lens has been identified using the CaII H \& K absorption lines and G-band. The source redshift is estimated based on  the  Lyman break which is somewhere in the Subaru IA624-band ($\lambda_{effective}$=6226~\AA, FWHM=299\AA), putting  constraints on  the source redshift: 3.9$<$z$_s<$4.1. From there, we find that for  $z_s$=3.96$\pm$0.02,  SiIV and CII  would align with spectral absorption features of the lens. This  estimation of the source redshift is used throughout the paper.\\
Moreover, with the  Low Resolution Imaging Spectrometer (LRIS), Lagattuta et al. (2010) measured the redshifts for seven lens galaxies and two sources. 

Finally, from the z-COSMOS catalog (Lilly et al. 2009) we have
retrieved the redshifts of seven strong lens candidates from our list
(based on the G-band and MgI absorption lines). Two of them,
re-observed with Keck/LRIS, benefit from a second redshift
determination from a higher signal-to-noise spectrum. Both redshifts
are in agreement.

A summary of the spectroscopic observations available for this sample is given in
Table~\ref{fig:obs}, while the lenses' redshifts are provided in
Table~\ref{treu}.

\begin{table}
\renewcommand{\arraystretch}{1.0}
\centering
\caption{Summary of the VLT (FORS1 and VIMOS) and Keck (DEIMOS and LRIS) observations: target name, instrument and exposure time (in ks). }
\label{fig:obs}
\begin{tabular}{c  c c   }
\hline
\hline
Name & Instr. & E.T. \\
          &                      &{\it ks}            \\
\hline
COSMOS~0012+2015 & FORS1& 1.8 \\
COSMOS~0013+2249 & FORS1 & 1.2 \\
COSMOS~0049+5128 & FORS1& 1.2 \\
COSMOS~5921+0638 & FORS1 &  1.8\\
\hline
COSMOS~0050+4901 & VIMOS     & 3.6    \\ 
COSMOS~0056+1226 & VIMOS &  3.6        \\
COSMOS~5947+4752 & VIMOS    &3.6        \\
J095930.93+023427.7& VIMOS & 3.6    \\
COSMOS~5857+5949 & VIMOS   &3.6        \\
COSMOS~0124+5121 & VIMOS   &3.6        \\
COSMOS~0227+0451 & VIMOS &3.6        \\
\hline
COSMOS~0018+3845 & DEIMOS & 12.6\\
\hline
COSMOS~0038+4123& LRIS & 3.6  \\
COSMOS~0050+4901& LRIS & 5.4  \\
COSMOS~0056+1226&LRIS &5.4 \\
COSMOS~0211+1139 &LRIS & 5.4  \\
COSMOS~0216+2955 &LRIS & 3.6  \\
COSMOS~0254+1430 & LRIS & 1.2  \\
J100140.12+020040.9   & LRIS & 5.4 \\
\hline
\end{tabular}
\tablefoot{
The dispersion is 5.52~\AA\, per pixel for FORS1/150I and  2.50~\AA\, per pixel for VIMOS/MR/OS-red. The VIMOS redshifts were obtained from the z-COSMOS follow-up of the field (Lilly et al. 2007).  The  Keck/DEIMOS observations were taken with the 830 line/mm grating, and using the OG550 blocking filter, leading to a dispersion of 0.47~\AA\, per pixel. The Keck/LRIS targets were simultaneously observed with both blue grating (300/5000, dispersion: 2.55~\AA\, per pixel) and red grating (600/7500, dispersion: 1.28 ~\AA\, per pixel) (see Lagattuta et al. 2010).
}
\end{table}

\begin{table}
\renewcommand{\arraystretch}{1.0}
\centering
\caption{Position and redshift of galaxies in the FORS1 field around COSMOS~0012+2015, COSMOS~0013+2249, and COSMOS~0049+5128. The error on the redshifts  is  $\pm$0.001. \label{tabexan}}
\begin{tabular}{c c  c}
\hline
\hline
RA & DEC& z\\
\hline
\multicolumn{3}{c}{COSMOS~0012+2015}\\
150.01554 & 2.3105342  & 0.492 \\
150.05263 &2.3377222   &0.378 \\
150.07391 &2.3256596  & 0.222 \\
150.07463 &2.3485102   &0.217 \\
150.07239 &2.3679651   &0.344  \\
150.08564 &2.3641732   &0.340  \\
150.10210 &2.3524922   &0.220 \\
150.11477 &2.3328146   &0.373  \\
\hline
\multicolumn{3}{c}{COSMOS~0013+2249}\\
150.10193 &2.3900368  &0.218 \\
150.09001 &2.3836205  &0.530  \\
150.08210   &2.3904710   &0.351  \\
150.07843 &2.3895686  &0.352  \\
150.07239   &2.3679506 & 0.347 \\
150.06510   &2.3855444 & 0.343\\
150.05808 &2.3804333  &0.346 \\
150.05235 &2.3837337  &0.350\\
150.03149 &2.3745112  &0.220\\
150.02774 &2.3738948  &0.222\\
150.00332 &2.4075260  & 0.347\\
\hline
\multicolumn{3}{c}{COSMOS~0049+5128}\\
150.23256 &1.8862956  &0.673 \\
150.23365 &1.8704970    &0.283  \\
150.21829 &1.8847159  &0.621 \\
150.20779 &1.8823028  &0.026 \\
150.20824 &1.8753444  &1.146 \\
150.20526 &1.8578028  &0.337  \\
150.19758 &1.8606685  &0.267  \\
150.18165 &1.8552841 & 0.168 \\
\hline
\end{tabular}
\end{table}

       \subsection{Improved photometric redshifts}
\begin{figure}[ht]
\begin{center}
\includegraphics[width=8cm]{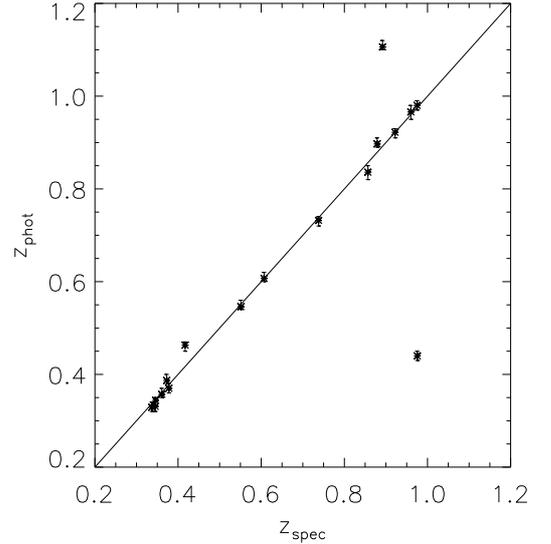}
\caption{ Photometric redshift, z$_{phot}$, versus spectroscopic
  redshift, z$_{spec}$, for 16 of the lens galaxy candidates. The
  solid line features z$_{phot}$=z$_{spec}$. Errors bars on the value
  of z$_{phot}$ relate to the 68\% confidence level.}\label{zz}
\end{center}
\end{figure}

Recently a new set of photometric redshifts for galaxies in the COSMOS
field has been derived by Ilbert et al. (2009), using the code {\it
  Le Phare} (S. Arnouts \& O. Ilbert\footnote{ Available at
  www.oamp.fr$/$people$/$arnouts$/$LE\_PHARE.html}) and the
photometric dataset by Capak et al. (2007a and 2009). Compared to the
previous set of COSMOS photometric redshifts by Mobasher et
al. (2007), which relied on {\it 8} broad bands, the main improvement
is the use  {\it 30} bands: the 8 broad bands plus 12 intermediate
bands, 2 narrow bands, 7 bands in the near-infrared and 1 band in the
ultra-violet. In Fig.~\ref{zz} we have displayed a comparison between
the new photometric redshifts and spectroscopic redshifts for the 17 lens
galaxies in our sample for which we have the two measurements. For most lens galaxies, the agreement is good within the 68\% confidence level error-bars measured for
the photometric redshifts. Thus, these new photometric redshifts for the lenses are reliable
estimates of their spectroscopic redshifts. We notice that three systems
show a strongly deviant photometric redshift (COSMOS~0254+1430: z$_{phot}$=0.46$\pm$0.01 and z$_{spec}$=0.417$\pm$0.001; J095930.93+023427.7: z$_{phot}$=1.10$^{+0.02}_{0.00}$ and z$_{spec}$=0.892$\pm$0.001; COSMOS~0018+3845: z$_{phot}$=0.44$\pm$0.01 and z$_{spec}$=0.9755$\pm$0.0003). In those three cases, the
photometry of the lens galaxy obtained from ground-based data is
contaminated by the close and bright images of the source, and the
photometric redshift determination is consequently biased. 

       \subsection{Stellar masses of the lens galaxies}\label{smlg}
\begin{figure}[ht]
\begin{center}
\includegraphics[width=9cm]{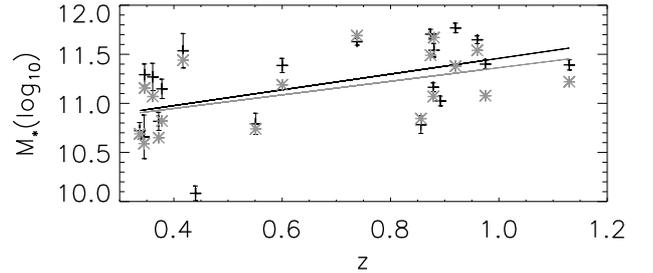}
\caption{The stellar masses (in M$_\odot$) of the lens galaxies as a
  function of redshift (spectroscopic when available, photometric
  otherwise). Black crosses: stellar masses inferred using the method
  described in Bundy et al. 2006. Grey stars: stellar masses inferred
  for the IRAC sources in Ilbert et al. 2010. Black and grey lines:
 the respective least square fits, with a slope of +0.8$\pm$0.3(black line) and  +0.7$\pm$0.3 (grey line).}\label{SMD}
\end{center}
\end{figure}

We have derived a first set of stellar masses using the Bayesian code
described in \citet{Bundy:2006}. In brief, a data couple, made of the
observed galaxy spectral energy distribution (SED) and its redshift,
is referenced to a grid of models constructed using the
\citet{bruzual:2003} synthesis code. The grid includes models that
vary in age, star formation history, dust content and metallicity.
At each grid point, the probability that the observed SED fits the
model is calculated: the corresponding stellar mass and the stellar
mass to K-band luminosity ratio are stored. By minimizing over all
parameters in the grid, the stellar mass probability distribution is
obtained. The median of this distribution is taken as the stellar mass
estimate, while the width corresponding to 68\% of the probability distribution encodes the error bar resulting from
degeneracies and uncertainties in the model parameter space. The final
error bar on the stellar mass also includes the K-band photometric
uncertainty as well as the expected error on the luminosity distance
that results from the uncertainty on the photometric redshift. The
stellar masses and uncertainties are provided in Table~\ref{treu} and
plotted in Fig.~\ref{SMD}.

From the ACS-F814W images (Koekemoer et al. 2007) we infer the
contamination due to the light coming from the background lensed
images in the stellar mass measurements.  In an aperture of 3\arcsec\,
radius, the lensed images are 3 to 7 magnitudes fainter than their
respective lensing galaxy: in all cases this contamination generates
an error on the stellar mass $\lesssim$ 0.01 dex.

In order to appreciate the reliability of our estimates of the stellar
mass (and uncertainties), we have retrieved among the 3.6$\mu$ IRAC catalog of the Spitzer-COSMOS survey (Sanders
et al. 2007) 18 of the 20 strong lenses; the other two lens galaxies were not detected with Spitzer. For these lenses we have compared our estimates of the
stellar masses to those inferred by Ilbert et al. (2010\footnote{The
  COSMOS catalogs can be found at
  http://irsa.ipac.caltech.edu/data/cosmos}): they are in reasonable
agreement (see Fig. \ref{SMD}). The typical difference between the two
mass estimators ($\sim$0.2 dex) is widely discussed in Ilbert et
al. (2010). Both distributions show a tendency of increasing lens
stellar mass with redshift. This feature will be discussed in more
detail in the discussion ($\S$~\ref{d1}).

\begin{table*}[ht]
\renewcommand{\arraystretch}{1.0}
\centering
\caption{\label{treu}  Summary of the properties of the lens galaxy environments (top to bottom: from low to high redshift). }
\begin{tabular}{ c  c  c  c  c  c  c c  r  }
\hline
\hline
Name & z &$i^+\pm\delta i^+$  &$\frac{\Sigma_{10}}{<\Sigma_{10}>_t}$& R$_{10}$&$\frac{D_{1}}{<D_{1}>_t}$ &N$_{1}$ & M$_\star$ & N$_{twin}$ \\
         & lens/source          &      &                               &    {\it kpc }         &              &&  {\it log10() M$_\odot$}  &         \\  
\hline
COSMOS~0049+5128 & 0.337$\pm$0.001             &  20.32$\pm$0.01  & 1.3$\pm$1.9 &         1013&1.0$\pm$0.9 & 10&10.72$\pm$0.08 &63\\
COSMOS~5947+4752 & 0.345$\pm$0.001             &  20.36$\pm$0.01  & 1.9$\pm$1.5 &         947&1.6$\pm$0.7 & 13&10.65$\pm$0.22 &59\\
COSMOS~0013+2249 & 0.346$\pm$0.001             &  19.63$\pm$0.01  &  6.2$\pm$0.2 &      602&2.3$\pm$0.5 &16&11.29$\pm$0.10 &23\\
COSMOS~0056+1226 & 0.361$\pm$0.001/0.808$\pm$0.001   & 19.77$\pm$0.01   & 1.4$\pm$1.2&           996&1.1$\pm$0.7 &11&11.26$\pm$0.13 &42\\
COSMOS~5857+5949 & 0.372$\pm$0.001             &  20.55$\pm$0.01  & 0.9$\pm$4.5  &      1348&0.4$\pm$2.2 & 5&10.81$\pm$0.09 &69\\
COSMOS~0012+2015 & 0.378$\pm$0.001             &  19.95$\pm$0.01  &   1.9$\pm$1.1 &       833&1.1$\pm$0.7 & 12&11.14$\pm$0.09 &40\\
COSMOS~0254+1430 & 0.417$\pm$0.001/0.779$\pm$0.001   &  19.76$\pm$0.01  & 0.5$\pm$3.5&           1767&0.6$\pm$1.1 &5 &11.53$\pm$0.17 &28\\
COSMOS~5921+0638 & 0.551$\pm$0.001/3.14$\pm$0.05   &  20.82$\pm$0.01  & 0.6$\pm$4.1&           1939&0.4$\pm$1.6 &3&10.79$\pm$0.10 &51\\ 
COSMOS~0216+2955 & 0.608$\pm$0.001             &  20.71$\pm$0.01  &   4.6$\pm$0.3    &     665&2.7$\pm$0.3 &19 &11.38$\pm$0.07 &29\\
COSMOS~0038+4133 & 0.738$\pm$0.001             &  21.03$\pm$0.01  &   5.5$\pm$0.9   &     585&1.9$\pm$0.5 & 17&11.62$\pm$0.03&30\\
COSMOS~0124+5121 & 0.856$\pm$0.001             &  22.61$\pm$0.02  & 2.9$\pm$1.0   &        777&2.1$\pm$0.4 & 15&10.78$\pm$0.08 &140\\
COSMOS~0047+5023 & 0.87$\pm$0.01               &  21.35$\pm$0.01  &  2.3$\pm$4.8   &      851&1.3$\pm$0.7 & 11&11.70$\pm$0.04 &22\\
J100140.12+020040.9  & 0.879$\pm$0.001            &  22.01$\pm$0.01  &  0.6$\pm$4.7  &     1931&0.3$\pm$2.6  & 2& 11.16$\pm$0.04 &94\\ 
COSMOS~5941+3628 & 0.88$\pm$0.01               &  21.60$\pm$0.01  &   2.5$\pm$0.4   &     946&2.0$\pm$0.3 & 12&11.54$\pm$0.06 &39\\
J095930.93+023427.7    & 0.892$\pm$0.001          &  21.98$\pm$0.01  & 1.2$\pm$2.4&              1131&0.8$\pm$0.9& 7  & 11.02$\pm$0.05 &92\\
COSMOS~0211+1139 &  0.92$\pm$0.01            &   21.65$\pm$0.01 &   1.8$\pm$0.6 &        948&1.6$\pm$0.4 & 13&11.76$\pm$0.04 &31\\
COSMOS~0050+4901 & 0.960$\pm$0.001             &   22.01$\pm$0.01  &0.8$\pm$4.9    &       1735&0.6$\pm$1.5 &4&11.64$\pm$0.03&43 \\
COSMOS~0227+0451 & 0.975$\pm$0.001             &   22.09$\pm$0.01  &   0.5$\pm$6.8  &     2149&1.0$\pm$0.9 &7&11.39$\pm$0.04&58 \\
COSMOS~0018+3845 & 0.9755$\pm$0.0003/3.96$\pm$0.02               &  23.43$\pm$0.03  &   1.0$\pm$1.3   &  1799     & 0.3$\pm1.2$ &5&10.73$\pm$0.06 &119\\
COSMOS~5914+1219 & 1.13$\pm$0.10               &   23.00$\pm$0.02  & 0.9$\pm$0.3     &    1832&0.7$\pm$1.2 &3 &11.39$\pm$0.05 &45\\
\hline
\end{tabular}
\tablefoot{
(1) Lens name. (2) Redshifts  of the lensing galaxies, followed by that of the source when known. (3) Subaru $i^+$ magnitude. (4) and (6) Number density ratios measured as explained in $\S$~\ref{voisin}. (5) Radius encompassing the 9$^{\rm th}$ closest neighbor.  (7) Total number of galaxies within 1Mpc from the lens. (8) Lens galaxy stellar mass. (9) Number of twins in the photometric catalog of Ilbert et al. (2010).
}
\end{table*}

\section{The environment of the lens galaxies}\label{densityratios}

 Having now more accurate redshift measurements for the lens galaxies and the galaxies in the COSMOS field, we can undertake a study of the local environments of lens galaxies. 
 Indeed, to understand whether high redshift lensing galaxies (z$\geq$0.3) are representative of the ETG population at their redshift, as the low redshift lens galaxies are 
 (T09, Auger 2008), we study their environment in comparison to the environment of a population of non-lensing ETGs. 
 Our first analysis of the lens environment with regard to LSS (Paper~II) led to the conclusion that  lens galaxies  are indeed evolving in the same environment than their 
 parent population. Lets now zoom into a more local environment for the lens, using neighbor density measurements  as  defined and used  by T09 for the SLACS 
 sample. 

\subsection{Projected number of neighbors}\label{voisin}

\begin{figure}[ht]
\begin{center}
\includegraphics[width=8cm]{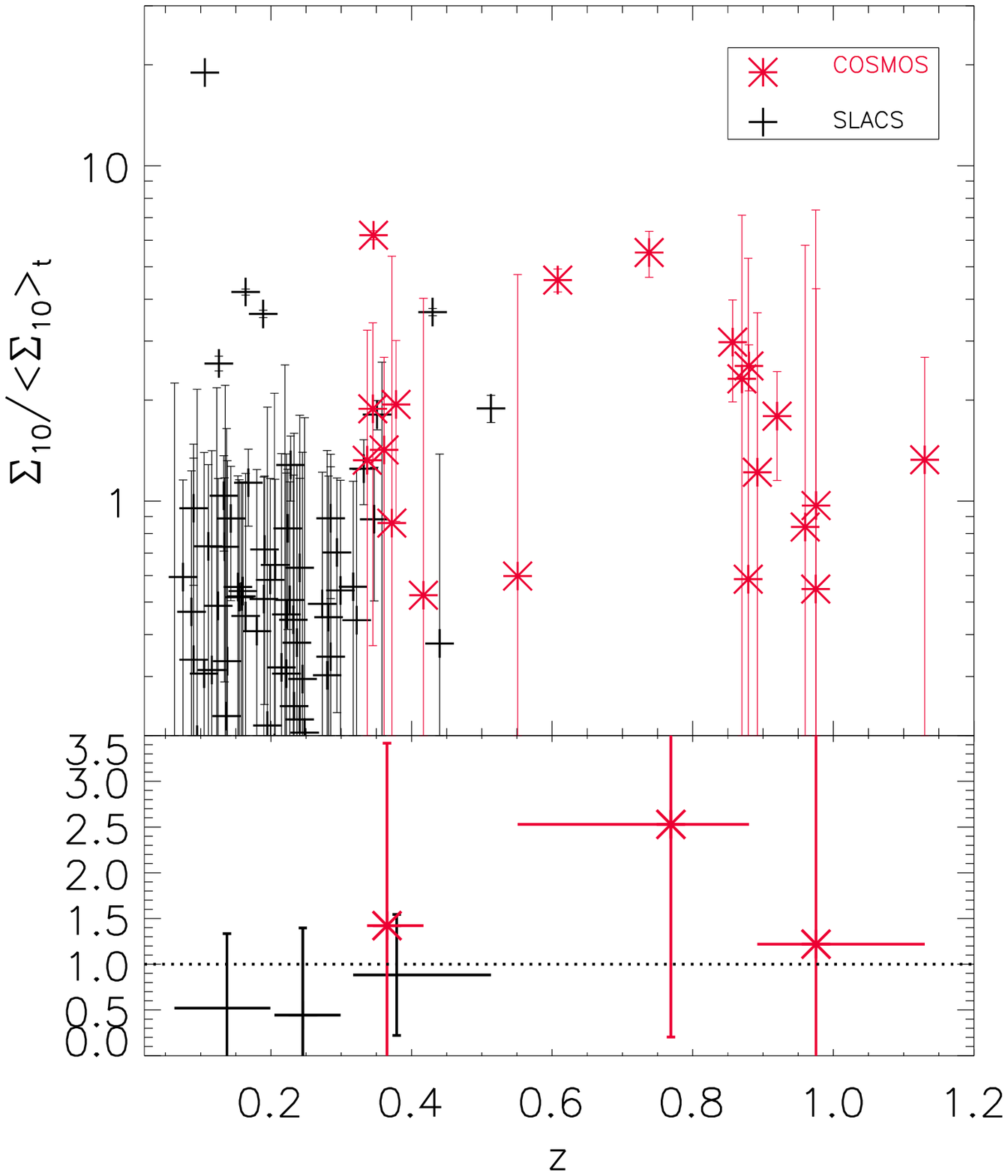}
\includegraphics[width=8cm]{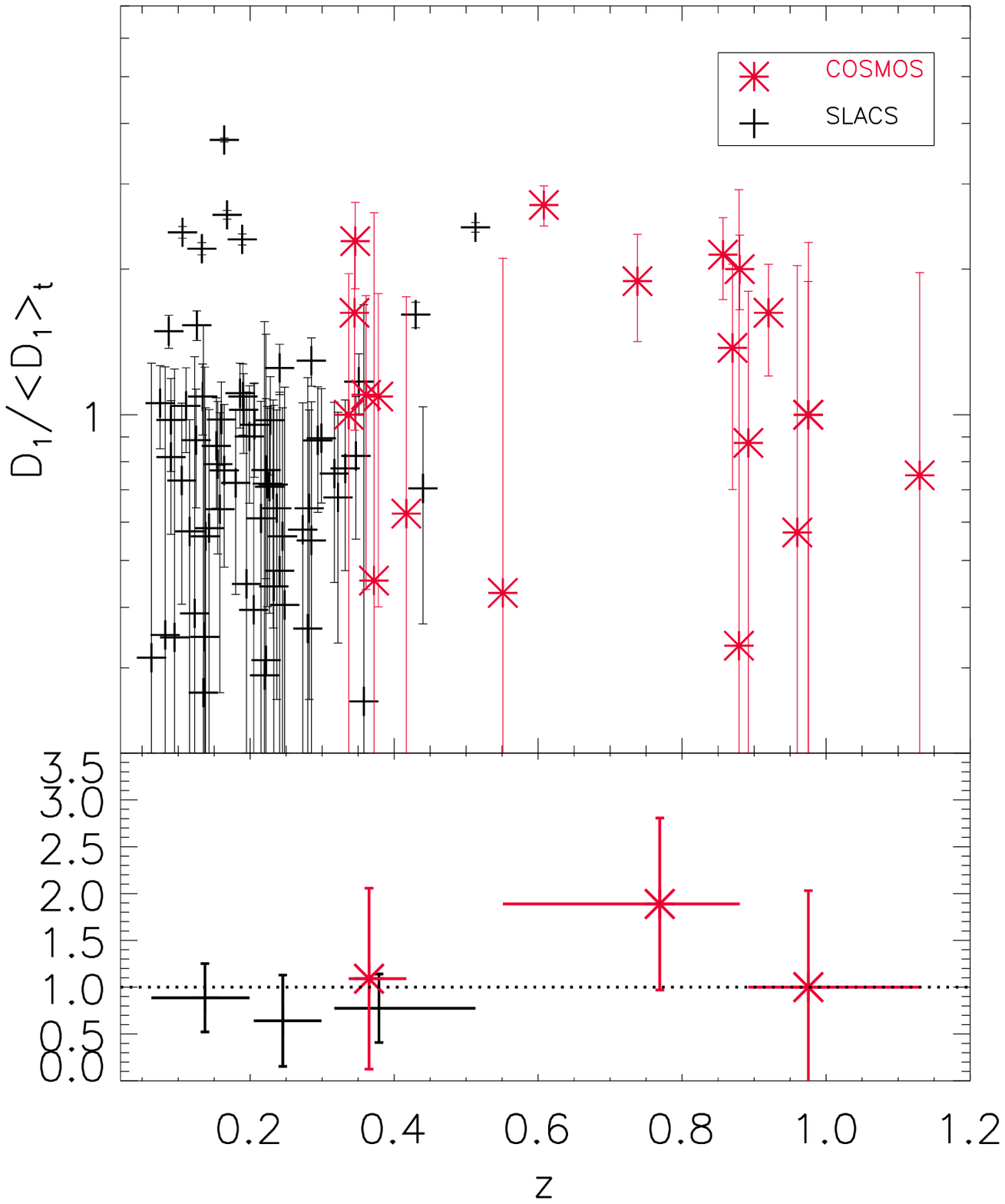}
\caption{The distribution of $\frac{\Sigma_{10}}{<\Sigma_{10}>_t}$ (top plot) and $\frac{D_1}{<D_{1}>_t}$ (bottom plot) as a function
  of the lens redshift, for the COSMOS lenses (red stars) and the  SLACS lenses (black crosses).  In each plot, the lower panel displays a re-binned version of the upper plot  using the  the median of the ratios in different bins (SLACS bins: z$<$0.2
  (32 galaxies), 0.2$\leq$z$<$0.3 (23 galaxies), 0.3$\leq$z$<$0.5 (9
  galaxies); COSMOS bins: 0.30$\leq$z$<$0.42 (7 galaxies),
  0.42$\leq$z$<$0.89 (7 galaxies), 0.89$\leq$z$<$1.30 (6 galaxies).  }\label{ratio_sig}
\end{center}
\end{figure}

\subsubsection{The neighbor number densities}

 In T09, the first estimator is the projected number density of galaxies,
$\Sigma_{10}$, inside a circle with radius R$_{10}$, equal to the
projected distance to the ninth closest neighbor of the lens. The
neighbor galaxies  are defined such as their magnitude is :
$i^+<i^+_{lens}$+3, and the upper or lower bound, z, of their photometric redshift is :
$z_{lens}-\delta z_{lens} < z < z_{lens} + \delta z_{lens}$ where
$\delta z_{lens}$ is a function of the lens magnitude and redshift,
and lies between 0.012 and 0.020 for our sample (see Ilbert et
al. 2009).  
For the two lenses with incorrect photometric redshifts, we
have used their spectroscopic redshift instead and for the search of
neighbors, the $\delta z_{lens} $ values associated with their
spectroscopic redshift in the photometric catalog; this ensures that we 
are  using the same search box for lens with spectroscopic and photometric redshifts.

The search for neighbors is performed in the photometric redshift
catalog of Ilbert et al. (2009) which contains about 1 million
objects.  The limiting magnitude chosen for the neighbors ensures that
the scaling in brightness (thus in mass) between a lens galaxy and
its neighbors is the same for all lenses, independently of its
magnitude. For the faintest lens in our sample, this upper limit
corresponds to the completeness limit of the COSMOS photometric
catalog (depth: i$^+\sim$~26.2, Capak et al. 2007a). The range chosen
for the redshift ensures that, given the error bars on the neighbor
photometric redshifts, they are indeed at the same redshift as
the lens galaxies (in the 68\% confidence limit error bars).

  As in T09, we also computed the projected number of galaxies, $D_1$, inside a
circle with radius 1~Mpc, at the lens redshift. To select the
neighbors, we have used the same limiting magnitudes and redshift
ranges as for the first estimator.

In order to get comparative measurements for a population of galaxies
which are not identified as lenses but have the same morphology (early
photometric type), similar brightness and redshift properties, we
define a control sample of ``twins'' for each lens galaxy. Those twins
have the upper or lower bounds, z$_{twin}$, of their photometric redshift such as: $z_{lens}-\delta z
<z_{twin}<z_{lens}+\delta z$ with $\delta z$=0.05, and the upper or lower bounds, ${\rm I}_{twin}$, of their magnitude such as :
 I$_{lens} - \delta {\rm I} < {\rm I}_{twin} < {\rm I}_{lens} - \delta{\rm I}$, with $\delta {\rm I}$=0.05~mag. The dispersion authorized around the value of the lens redshift and lens magnitude ensures that the sample of twins is large enough for the results to be statistically reliable (between 22 and 140 twins, see Table~\ref{treu}).  In addition, the range in redshift is small enough to avoid possible biases due to galaxy evolution and the range in  magnitude ensures that the flux difference between a lens and its twins remains lower than 5\%. We then measure the
average number density of neighbors $<\Sigma_{10}>_t$ and the average
projected number of galaxies in a circle with radius 1~Mpc: $<D_{1}>_t$
for the control ``twin'' sample. 

 We notice that, for an obvious lack of spectroscopic information, the twins cannot be selected according to their velocity dispersion as it is the case in the analysis by T09.  
 Yet our match in magnitude and redshift mimics, up to a certain level of accuracy,  a match in mass between the lens galaxies and their twins.

\subsubsection{Error measurements on the neighbor number densities}

In building the photometric catalog of galaxies, masks were necessary to hide bright objects disturbing the galaxy extraction
procedure (see Capak et al. 2007a). From these masks we have measured, at each lens and twin location,
the surface that was hidden for the extraction. The hidden surface is typically of a few percents of the total surface used to determine   $\Sigma_{10}$ or  $D_1$. It only gives a positive contribution to the total error budget  on the neighbor density estimates: indeed, the surface covered by the circle of radius R$_{10}$ would be smaller if the ninth closest galaxy was behind the mask, and if one or more galaxies are hidden behind the mask,  the number of galaxies in a circle of radius $D_1$ would be larger. Therefore, for a  fixed number of galaxies, the surface has to be smaller.  
 In summary, for a lens galaxy, the total error bar on the neighbor
galaxy density is the quadratic sum of the Poisson error and the error due to
the use of masks on the surface encompassing the neighbors. For its
average twin, the error bar is the quadratic sum of the Poisson error, the
error due to the use of masks and the dispersion for the population
of twins. The Poisson error largely dominates the error budget for the
lens neighbor number densities, while the dispersion largely
dominates the error budget for the twins.

\subsubsection{Results}
The values of $\Sigma_{10}$, $D_1$, $<\Sigma_{10}>_t$ and $<D_{1}>_t$ depend on the redshift bin and magnitude limit. 
To circumvent these effects, we only interpret the results in term of ratios: 
$\frac{\Sigma_{10}}{<\Sigma_{10}>_t}$ and $\frac{D_1}{<D_{1}>_t}$. 
These ratios quantify the richness of the environment
of the lens galaxies in comparison to the environment of their
respective twins. They are provided in Table~\ref{treu}. The average
distance encompassing the 9 closest galaxies to the lens is $<$R$_{10}>$=1.2~Mpc with a 1$\sigma$ standard deviation of 0.5~Mpc,
hence comparable to  the radius of 1~Mpc used to calculate $D_1$. In Fig.~\ref{ratio_sig}, we have displayed the
ratios $\frac{\Sigma_{10}}{<\Sigma_{10}>_t}$ and $\frac{D_1}{<D_{1}>_t}$ for the
COSMOS lenses and for the SLACS lenses (using data from T09), as a function of redshift. {\it A-priori}, the two
distributions cannot be directly compared as they may have different
normalization factors (due to their different lens selection function,
source redshift distribution and survey sensitivity). Yet, we notice
that in the redshift range common to the SLACS and COSMOS samples
(z$\sim$[0.33,0.50]), the levels of the distributions for
$\frac{\Sigma_{10}}{<\Sigma_{10}>_t}$ and $\frac{D_1}{<D_{1}>_t}$ are similar,
suggesting that the normalizations are not drastically
different. However, in order to avoid any misinterpretation due to
intrinsic sample differences, the SLACS sample and the COSMOS sample
distributions are analyzed separately.

The number density ratios displayed in Fig.~\ref{ratio_sig} indicate
whether the lens galaxies reside in a typical environment in
comparison to the twin galaxy population. Would this be true, the
number density ratios should be of the order of 1 at every redshift. 
This is indeed the
case in average for the lenses in the redshift range studied here.\\

\subsection{X-ray and optically detected galaxy groups and clusters}\label{optix}
The validity of the estimators used to characterize the environments of lens galaxies in  $\S$~\ref{voisin}  can be checked using the distribution of groups and clusters in the field based on X-ray and spectroscopic data analysis. Indeed, we can access two supplementary pieces of information: (a) the
distribution of X-ray emitting gas in the COSMOS field (XMM-Newton
survey: Hasinger et al. 2007 and C-COSMOS with Chandra: Elvis et
al. 2009), which traces galaxy groups and clusters in the field
(Finoguenov et al. 2007 and in prep.), and (b) the
identification of optical groups or clusters (Knobel et al. 2009) from
the z-COSMOS spectroscopic program (Lilly et al. 2007). 

Moreover, as the presence of groups in the direction to the source generates a different lensing potential in comparison to that of a single lens galaxy, it is important, in preparation of the mass modeling in $\S$~\ref{model} to know: 1) If the lens galaxies are group members. If this is indeed the case,  the group has to be modeled as an individual potential in the lens modeling.  2) What contribution in terms of shear is produced by the groups and clusters towards and around the lens galaxy. Most of the time this contribution is simply represented by an ''external shear" whose parameters are optimized while adjusting  the  lens model. In this paper, we rather intent to measure independently the external shear using the rich dataset that covers the COSMOS field.
      
\begin{table*}[ht]
\renewcommand{\arraystretch}{1.0}
\centering
\caption{\label{knobfig}  Summary of the contribution of the environment to the lens galaxies. }
\begin{tabular}{  c  c   c  c c r c  c  }
\hline
\hline
Name & P  &N&$\kappa$& $\gamma$ &$\rm{PA}_\gamma$ & Cluster  & Dist.\\
    &   &     &   & & {\it deg}          & ID (z$\pm\delta$z, log10(M$_{200}$), c, r$_{200}$) &{\it kpc}\\
\hline
COSMOS~0049+5128 & 0.33& 13&  0.006$\pm$0.001$^{+0.000}_{-0.001}$  & 0.008$\pm$0.003$^{+0.001}_{-0.001}$ & -38$\pm$12$^{+0}_{-1}$ &  \_&  \_ \\
COSMOS~5947+4752 & 0.04& 3& 0.016$\pm$0.003$^{+0.006}_{-0.001}$  & 0.021$\pm$0.005$^{+0.008}_{-0.006}$ & 0$\pm$2$^{+0}_{-3}$  & \_&  \_\\                 
COSMOS~0013+2249  & 0.33& 26& 0.023$\pm$0.003$^{+0.000}_{-0.008}$  & 0.025$\pm$0.005$^{+0.001}_{-0.002}$  &  8$\pm$10$^{+1}_{-1}$   &\#173 (0.348$\pm$0.005,13.44,4.18,0.55)&0.5 \\
COSMOS~0056+1226 & 0.76 & 29& 0.011$\pm$0.003$^{+0.000}_{-0.000}$  & 0.003$\pm$0.004$^{+0.000}_{-0.000}$  & -18$\pm$30$^{+0}_{-0}$ &\#133 (0.360$\pm$0.005,13.28,4.32,0.48)&0.3\\
COSMOS~5857+5949 & 0.58& 18& 0.019$\pm$0.005$^{+0.002}_{-0.003}$   & 0.033$\pm$0.010$^{+0.005}_{-0.004}$  & -41$\pm$8$^{+0}_{-0}$ &\_ &  \_\\
COSMOS~0012+2015  & 0.39 & 31& 0.019$\pm$0.002$^{+0.000}_{-0.003}$ & 0.009$\pm$0.006$^{+0.000}_{-0.001}$  & -24$\pm$19$^{+5}_{-0}$ & \_ &  \_\\
COSMOS~0254+1430 & 0   & 2 & 0.002$\pm$0.000$^{+0.000}_{-0.000}$    & 0.006$\pm$0.000$^{+0.000}_{-0.000}$ & -31$\pm$3$^{+0}_{-0}$ &\_&  \_\\
COSMOS~5921+0638 & 0.61& 24& 0.019$\pm$0.004$^{+0.000}_{-0.000}$   & 0.009$\pm$0.012$^{+0.000}_{-0.000}$ & -37$\pm$31$^{+0}_{-0}$&\_ &  \_\\  
COSMOS~0216+2955 & 0.06 & 5  & 0.025$\pm$0.003$^{+0.005}_{-0.005}$& 0.024$\pm$0.004$^{+0.000}_{-0.003}$ & 36$\pm$5$^{+3}_{-4}$ &\#221 (0.600$\pm$0.005,13.56,3.82,0.55)&0.2\\ 
COSMOS~0038+4133  & 0.12   & 4& 0.004$\pm$0.000$^{+0.001}_{-0.001}$& 0.014$\pm$0.002$^{+0.004}_{-0.003}$ & -4$\pm$4$^{+2}_{-0}$   &\_&  \_\\
COSMOS~0124+5121 & 0.20 & 2& 0.002$\pm$0.000$^{+0.001}_{-0.001}$   & 0.007$\pm$0.002$^{+0.000}_{-0.000}$ & 9$\pm$3$^{+2}_{-0}$  &\_  &  \_\\
COSMOS~0047+5023 & 0.06 & 4  & 0.012$\pm$0.002$^{+0.002}_{-0.002}$ & 0.016$\pm$0.003$^{+0.004}_{-0.003}$ & -26$\pm$6$^{+0}_{-0}$ &\_&  \_\\
J100140.12+020040.9   & 0.20  & 3 & 0.006$\pm$0.001$^{+0.001}_{-0.000}$ & 0.009$\pm$0.002$^{+0.001}_{-0.001}$ & -28$\pm$7$^{+0}_{-0}$&\_&  \_\\
COSMOS~5941+3628 & 0.07 & 7& 0.024$\pm$0.003$^{+0.004}_{-0.005}$   & 0.007$\pm$0.004$^{+0.010}_{-0.004}$& 22$\pm$15$^{+9}_{-1}$  & \_&  \_\\
J095930.93+023427.7 & 0.27& 4 & 0.006$\pm$0.001$^{+0.002}_{-0.001}$    & 0.016$\pm$0.004$^{+0.002}_{-0.001}$ & -45$\pm$1   &\#288 (0.697$\pm$0.006,13.63,3.73,0.56)&102 \\
COSMOS~0211+1139 & 0  & 3   & 0.004$\pm$0.000$^{+0.001}_{-0.001}$    & 0.022$\pm$0.005$^{+0.006}_{-0.005}$ & 43$\pm$3$^{+0}_{-1}$  &\_&  \_\\ 
COSMOS~0050+4901 & 0.25  & 2  & 0.001$\pm$0.000$^{+0.001}_{-0.000}$ & 0.006$\pm$0.001$^{+0.001}_{-0.001}$ & -16$\pm$4$^{+5}_{-3}$ &\_&  \_\\             
COSMOS~0227+0451 & 0   & 4 & 0.002$\pm$0.000$^{+0.001}_{-0.001}$     & 0.005$\pm$0.001$^{+0.002}_{-0.002}$ & -19$\pm$6$^{+4}_{-9}$  &\# 101 (0.98$\pm$0.01,13.70,3.65,0.53)&0.5\\        
COSMOS~0018+3845  & 0.38 & 15 & 0.037$\pm$0.007$^{+0.000}_{-0.000}$ & 0.082$\pm$0.016$^{+0.000}_{-0.000}$  & 13$\pm$4$^{+0}_{-0}$ & \_&  \_\\             
COSMOS~5914+1219 & 0.25 & 17 & 0.001$\pm$0.000$^{+0.000}_{-0.000}$   & 0.006$\pm$0.001$^{+0.000}_{-0.000}$& -19$\pm$5$^{+1}_{-0}$  &\_ &  \_\\
\hline
\end{tabular}
\tablefoot{(1) Lens name. (2) Probability that galaxies in a 2\arcmin$\times$2\arcmin\ field around the lens galaxy are included in the z-COSMOS catalog.  (3) The number of groups involved in the calculation of the external shear (within 5\arcmin\, projected radius). (4, 5, 6) Convergence, shear strength and shear direction due to the groups and clusters around the lens. The first set of error comes from the uncertainty on the group mass. The second set of error are due to the uncertainty on the source redshift, chosen to be $\pm$0.5 for sources with unknown redshifts and the error on the source redshift estimation from Table~\ref{treu} when it has been measured. (7) The ID of the group or cluster with an impact parameter smaller than its radius (r$_{200}$) in Leauthaud et al. (2010). In parenthesis are its redshift, mass (in solar mass), concentration and radius (in Mpc). The error on the cluster mass is 40\% of the total mass (Vikhlinin et al. 2009). (8) Projected distance in kpc between the lens and the group/cluster centre (at the lens redshift). }

\end{table*}

In the framework of our lensing analysis, each of the group catalogs
(X-ray selected versus spectroscopically selected) has certain
advantages and disadvantages. The entire COSMOS region has been mapped
through 54 overlapping XMM-Newton pointings while additional Chandra
observations have mapped the central region (0.9 square degrees). A
composite XMM-Newton and Chandra mosaic has been used to detect and
measure the fluxes of groups and clusters to a 4$\sigma$ detection
limit of $1.0 \times 10^{-15}$ erg cm$^{-2}$ s$^{-1}$ over 96\% of the
ACS field. The general data reduction process can be found in
Finoguenov et al. (2007, 2009) and the cluster catalog and weak-lensing analysis are presented in
Leauthaud et al. (2010). Therefore, the catalog of X-ray selected
groups and clusters is homogeneous and covers the entire COSMOS field.
However, it is affected by the sensitivity threshold of the X-ray
survey, which is different from the sensitivity cut of the optical
dataset. Therefore, groups of modest mass could be missing from
the X-ray selected catalog, especially around galaxies at high
redshift.

Conversely, the catalog of
optically selected groups in COSMOS has the advantage of spanning the
same brightness range as the optical imaging dataset from which the
strong lens candidates and the arcs have been extracted. Even though
the faintest galaxies seen on deep images will remain undetected in
the shallower spectroscopic z-COSMOS survey, groups can still be traced
and assessed from their central brightest galaxies. Then, the
parameters of the galaxy groups can be derived by comparing these
detections to a catalog of mock-groups subject to the same detection
criteria (see Knobel et al. 2009 for an extensive description of the
group catalog). The first data release of the z-COSMOS
survey (Lilly et al. 2007) covers a fraction of the full COSMOS
field, which leaves some of our COSMOS lens candidates outside the
coverage of the group catalog.

Let us establish now whether the lens galaxies are group or cluster members ($\S$~\ref{lggm}) and measure the external shear contribution of the groups and clusters at the lens galaxy position ($\S$~\ref{lenscont}).

\subsubsection{Lens galaxy group members}\label{lggm}

We cross-correlate the X-ray
cluster catalog (Leauthaud et al. 2010) with the sample of strong lenses.  We define as a group member, a lens galaxy located  within r$_{200}$ (radius where the matter density is 200 times the critical density) of the group center, and with redshift  identical to the group redshift (within error bars). We have found  4 matches and there is a galaxy cluster detected at a lower-redshift in direction to J095930.93+023427.7 (see Table~\ref{knobfig}).

Using the group catalog built from the optical dataset, 
we do
not identify any new groups associated in redshift and observational plane with lens galaxies than those already identified
using the X-ray catalog. 
 In Table~\ref{knobfig} we give the
probability that galaxies in a 2\arcmin$\times$2\arcmin\ field around the lens galaxy are included in the z-COSMOS catalog. This figure
establishes roughly the ``completeness'' of the survey at every lens
location and tells us if the z-COSMOS group catalog can be used to
characterize the environment of the lens.

\subsubsection{Comparison:  galaxy group members and  projected number density of neighbors}\label{gvp}

 In principle, lens galaxies which are group members are expected to have large neighbor number densities. This should be visible in the ratio  measured in $\S$~\ref{voisin}. This is indeed the case with the first estimator for the lens in COSMOS~0013+2249 ($\frac{\Sigma_{10}}{<\Sigma_{10}>_t}$=6.2$\pm$0.2) and  COSMOS~0216+2955 ($\frac{\Sigma_{10}}{<\Sigma_{10}>_t}$=4.6$\pm$0.3), and it is unclear for COSMOS~0056+1226 and COSMOS~0227+0451 because of the large error bar on the measured density ratios. With the second estimator, the ratio are in average lower than with $\frac{\Sigma_{10}}{<\Sigma_{10}>_t}$, and it is therefore more difficult to identify correctly an over-dense field.
 On the contrary, COSMOS~0038+4133 shows a large density ratio ($\frac{\Sigma_{10}}{<\Sigma_{10}>_t}$=5.5$\pm$0.9) without being associated to any known galaxy group in X-ray or optic.

\subsubsection{The external shear due to clusters and groups}\label{lenscont}

 We estimate the shear and
convergence produced by the groups (either X-ray or optical) detected
around the  lens galaxies  in order to characterize the contribution of the
environment to the total lens potential in future lens models ($\S$~\ref{model}). To do so we followed
the method described in Paper~III. In short, we compute
the convergence and shear produced by all groups closer in projection
than a given radius (5\arcmin, as in Paper~III), and at any redshift
up to that of the source. We assume that the mass profile of every group
follows a truncated isothermal sphere (TIS). The choice
of this profile is motivated by the fact that the shear and
convergence produced by an isothermal sphere are easy to compute
(Keeton 2003, Momcheva et al. 2006). Moreover, as we will consider
only groups and clusters which do not cross the line-of-sight to the
source, estimations of their total masses rather than of their mass
distributions are sufficient for the accuracy of the result. We
selected a truncated profile to avoid to give unrealistically too much weight to the
most distant groups.
 
The  three-dimensional density distribution of the TIS can be written:
\begin{equation}
\rho \propto \frac{1}{r^2} \frac{1}{r^2+r_{c}^2}
\end{equation}

\noindent where $r_c$ is the truncation radius. The convergence, $\kappa$, and shear, $\gamma$, produced  by this profile are respectively:

\begin{equation}
 \kappa=\frac{\tilde{b}}{2} \left(\frac{1}{\sqrt{r^2}}-\frac{1}{\sqrt{r_{c}^2+r^2}}\right) 
 \end{equation}
 \begin{equation}
 \gamma=\frac{\tilde{b}}{2} \left[\frac{1}{\sqrt{r^2}}+\frac{1}{\sqrt{r_{c}^2+r^2}}-\frac{2r_{c}}{r^2}\left(\frac{\sqrt{r_{c}^2+r^2}}{r_{c}}-1\right)\right].
\end{equation}
\noindent where $\tilde{b}$ is the impact parameter of the TIS, which relates to the Singular Isothermal Sphere (SIS) impact parameter $b_{SIS}$ as:

\begin{equation}
\frac{b_{SIS}}{\tilde{b}}=1+\frac{r_c}{b_{SIS}}-\sqrt{1+\left(\frac{r_c}{b_{SIS}}\right)^2}.
\end{equation}
In the limit where $r_c\rightarrow \infty$ these quantities match those of the SIS.

To compute the external shear produced by the groups and clusters we proceed as follows.
Whenever  P$<$0.3 in Table~\ref{knobfig} (column 2), we only consider the catalog of X-ray
detected groups and clusters. The groups mass potentials are
modeled as TIS using the value M$_{200}$  and r$_{200}$ (used as the truncature radius). The error on the X-ray mass comes from the scatter in the relation used to derive the mass from the luminosity (Leautaud et al. 2010): it is  of the order of $\pm$20\% of M$_{200}$  (Vikhlinin et al. 2009). Whenever P$\ge$0.3 in Table~\ref{knobfig} (column 2), we also consider the optical group
catalog in addition to the X-ray group catalog. We first correlate the optical and X-ray group catalogs to
identify and remove optical groups that might already be accounted as 
an X-ray group. Then, we model as TIS the remaining optical groups using the
``mock'' virial mass, M$_{vir}$, and the ``mock'' virial radius, r$_{vir}$  (used as the truncature radius). The upper and lower error on the virial mass are $^{+100\%}_{-50\%}$ of M$_{vir}$ (Knobel et
al. 2009). These ``mock'' 
quantities are the theoretical values associated with the detected
groups when subjecting a mock sample of groups to the same selection
function and same survey criteria as the observations (see Knobel et
al. 2009 for details).

The external shear and associated convergence are calculated for each group individually. They are then re-scaled to the redshift of the lens  using the scaling relation given in Momcheva et al. (2006). For a given lens, external shear and convergence are  summed up
following the procedure described in Keeton (2003) and Momcheva et
al. (2006). The results are summarized in Table~\ref{knobfig}. A first set of errors on the shear and on the convergence results from the propagation of the group mass errors. 

As mentioned already, in the special case where a galaxy group or
cluster has an impact parameter smaller than $r_{200}$ or $r_{vir}$ (in
5 cases, see Table~\ref{knobfig}), the shear and convergence
calculated under this simple approximation are incorrect (Keeton
2003). Hence, we have systematically removed these groups when computing the external shear. Instead, they will have to be accounted for additional lens potential when performing the lens modeling.

Regarding the source redshift, it is either known spectroscopically
for some lenses (see Table~\ref{treu}), or assumed to be at z$_s$=2
for lenses with z$_l$$<$1, or at z$_s$=3 for lenses with
z$_l$$>$1. The convergence and shear contributions depend on the
number of groups taken into account, which in turn may depend on the
source redshift and on the cut-radius considered.  Hence, an error on the source
redshift generates an error on the external shear parameter. We have estimated this error assuming an uncertainty $\delta z$=$\pm$0.5 on the source redshift when unknown, and using the error on the source redshift in Table~\ref{treu} when measured (see Table~\ref{knobfig}).
To estimate the error introduced by an arbitrary
cut-radius at 5\arcmin, we have also probed two other radii (7\arcmin\, and
10\arcmin). We find that the incidence of the radius selection on the
final lensing contribution is $\delta\kappa\sim\delta\gamma\sim$0.001
and is negligible on the orientation of the shear ( $\delta \rm{PA}_\gamma<$0.5$^\circ$).  
Then we calculate the total error on the external shear 
as the sum in quadrature of the error produced by the group mass uncertainty,  the source redshift uncertainty and the error produced by the choice of aperture.  

We have also analyzed the error on the external shear parameters generated when using  a single catalog of groups (X-ray) instead of the combination of the two catalogs (X-ray and optical).  The comparison is possible for 7 systems. We notice that, while the convergence is different when using a single catalog instead of two, the external shear strength and direction are in agreement within their respective error bars. This  means that our calculation realistically associates the largest source of shear with the more massive groups and clusters. It also means that  the probable incompleteness of our catalogs in low mass groups has a minimal impact on the shear measurement: what matters is the completeness of the catalog in large mass groups and clusters. However,  the technique is limited for high redshift sources, as there is no X-ray cluster detected above z=1.3 in the COSMOS field.

\section{The strong lens modeling}\label{model}
 In this section, we focus on the 12 lens galaxies offering the largest number of observational constraints: the triple and quadruple image systems and the Einstein rings.

\subsection{Lens galaxy light profiles}\label{LPF}

\begin{center}
\begin{table*}
\centering
\caption{\label{priors} Lens galaxy luminous profile.}
\begin{tabular}{ c c c c c c }
\hline
\hline
Name  & \multicolumn{5}{c}{Sersic parameters}\\
               & $\epsilon$&PA&$R_{e}$ &$R_{e}$&n  \\
    &                     &{\it deg}& {\it kpc} &{\it \arcsec}&  \\
    \hline
    COSMOS~0049+5128    & 0.22$^{+0.00}_{-0.01}$   & -25$^{+1}_{-1}$ &5.15$^{+0.04}_{-0.03}$ &1.10$^{+0.01}_{-0.01}$  &1.19$^{+0.01}_{-0.02}$        \\
  COSMOS~5947+4752  & 0.05$^{+0.15}_{-0.05}$   &  4$^{+7}_{-16}$   &2.49$^{+1.53}_{-0.54}$  & 0.52$^{+0.32}_{-0.11}$   & 1.38$^{+0.53}_{-0.36}$ \\ 
    COSMOS~5921+0638    &0.14$^{+0.06}_{-0.03}$     &   97$^{+18}_{-8}$  &2.88$^{+3.37}_{-0.11}$ &0.46$^{+0.54}_{-0.02}$     &1.00$^{+1.35}_{-0.04}$  \\   
    COSMOS~0038+4133 & 0.25$^{+0.08}_{-0.09}$   &  -3$^{+26}_{-4}$ &5.25$^{+0.81}_{-2.37}$&  0.74$^{+0.11}_{-0.33}$  & 4.30$^{+0.50}_{-1.00}$   \\
    COSMOS~0124+5121   & 0.23$^{+0.16}_{-0.03}$   & -52$^{+28}_{-4}$&1.82$^{+0.15}_{-0.17}$ &  0.24$^{+0.02}_{-0.02}$ & 1.89$^{+0.29}_{-0.14}$  \\ 
    COSMOS~0047+5023    &0.19$^{+0.01}_{-0.01}$  & 33$^{+2}_{-2}$ &5.40$^{+0.10}_{-0.08}$& 0.72$^{+0.01}_{-0.01}$ &1.25$^{+0.03}_{-0.03}$  \\
    J100140.12+020040.9  & 0.16$^{+0.06}_{-0.02}$ &   21$^{+8}_{-6}$&2.42$^{+0.22}_{-0.52}$ & 0.32$^{+0.03}_{-0.07}$ &2.59$^{+0.13}_{-0.23}$   \\
    COSMOS~5941+3628   & 0.24$^{+0.10}_{-0.16}$    & -5$^{+22}_{-17}$  & 5.89$^{+0.22}_{-0.24}$ & 0.78$^{+0.03}_{-0.03}$ &1.18$^{+0.36}_{-0.10}$   \\  
    J095930.93+023427.7     & 0.21$^{+0.05}_{-0.05}$    & -77$^{+6}_{-7}$&1.62$^{+0.50}_{-0.68}$ & 0.21$^{+0.07}_{-0.09}$   &1.90$^{+0.34}_{-0.79}$   \\
    COSMOS~0050+4901   & 0.30$^{+0.06}_{-0.03}$   &27$^{+3}_{-10}$   &2.85$^{+0.47}_{-1.00}$  & 0.37$^{+0.06}_{-0.13}$ &5.58$^{+0.19}_{-0.13}$     \\
    COSMOS~0018+3845 & 0.22$^{+0.00}_{-0.01}$    &  -22$^{+1}_{-1}$ &2.32$^{+0.54}_{-1.31}$  &  0.30$^{+0.07}_{-0.17}$  & 5.60$^{+0.19}_{-0.13}$      \\ 
 COSMOS~5914+1219    &0.13$^{+0.02}_{-0.10}$    &14$^{+48}_{-2}$   &2.21$^{+0.06}_{-1.15}$ & 0.27$^{+0.01}_{-0.13}$ &1.40$^{+2.28}_{-0.06}$     \\
    \hline 
    \end{tabular}
    \tablefoot{(1) Lens name.  Parameters of the Sersic light profile fit: (2) ellipticity, (3) position angle, (4) effective radius in kpc and (5) in arc-seconds and (6) index, with associated error bars (from GIM2D, 68\% confidence limit) .  }
\end{table*}
\end{center}

\begin{table}
\centering
\caption{Position of the lensing galaxy (in degree) and relative position (in arc-second) of the images used as constraints for the lens mass model. \label{tabexappen2}}
\begin{tabular}{c c  c  }
\hline
\hline
Lens &RA&DEC\\
{\it  Image}& {\it  $\delta$RA }&{\it $\delta$DEC}\\
\hline
COSMOS~0049+5128& 150.2052807& 1.8578028\\
{\it  A1} &{\it 2.05}&{\it  0 } \\
{\it A2} &{\it 0 }&{\it 2.05  }\\
{\it A3 }&{\it 0} &{\it -2.05 } \\
{\it A4 }&{\it -2.05} &{\it 0} \\
\hline
COSMOS~5947+4752& 149.94955 &2.7979502\\
{\it A1}& {\it 2.28} &{\it 0 }\\
{\it A2} &{\it -2.28}&{\it  0} \\ 
{\it A3 }&{\it 0} &{\it 2.28 }\\ 
{\it A4 }&{\it 0 }&{\it -2.28} \\ 
\hline
COSMOS~0038+4133&150.1595 &2.6927351\\
{\it A1} &   {\it -0.43  }& {\it  -0.33}  \\
{\it  A2}&   {\it 0.00  } & {\it -0.62  }  \\
{\it A3  }  & {\it 0.04 }&  {\it  -0.61 }\\
{\it  A4  }&  {\it  0.22 }  & {\it  0.71 }  \\
\hline
COSMOS~0124+5121&150.3522 &1.8558994\\
{\it  A1} &{\it 0}& {\it 0.89}\\
{\it A2 }&{\it 0} &{\it -0.89}\\
{\it A3 }&{\it 0.89}& {\it 0 }\\
{\it A4 }&{\it -0.89} &{\it 0}\\
\hline
COSMOS~0047+5023& 150.19858 & 1.8397919\\
{\it  A1}& {\it -1.42}&{\it  -1.15}\\
{\it A2 }&{\it 0.10 }& {\it -1.56} \\
{\it A3 }&{\it 2.08 }&{\it -0.36}  \\
{\it  A4}&{\it  -0.05}&{\it 0.90}  \\
\hline
COSMOS~5941+3628& 149.92209& 2.6080412 \\
{\it  A1 }&{\it 0.} &{\it 1.24}  \\
{\it A2 }&{\it 0.}& {\it -1.24 } \\
{\it A3 }&{\it -1.24}&{\it 0  }\\
{\it  A4 }&{\it 1.24}& {\it 0  } \\
\hline
COSMOS~0050+4901&150.21104 &2.8171935\\
{\it A1} &{\it 1.15 }&{\it 0.69 }\\
{\it A2}&{\it- 1.53 }&{\it 1.10 } \\
{\it A3}& {\it -1.53}&{\it -0.80 } \\
{\it  A4}&{\it 0.27} &{\it 1.88  }\\
\hline
COSMOS~0018+3845 &150.07666 & 2.6458333\\
{\it A1}&{\it  -0.39}& {\it -1.15}\\
{\it A2} &{\it -1.22} &{\it -0.33 }\\
{\it A3 }&{\it 1.36}&{\it 0.25}\\
\hline
COSMOS~5914+1219&149.81142 & 2.2054236\\
{\it  A1}&{\it -0.93 }& {\it -0.45}\\
{\it A2 }&{\it 1.26 }& {\it 1.43 }\\
{\it A3}& {\it 1.81}&{\it 0.55 } \\
\hline
\end{tabular}
\end{table}

We have re-computed the 2-dimensional fit of the galaxy surface brightness 
distribution of the COSMOS lenses using GIM2D (Simard 1998, Marleau \& Simard 1998) in order to include error bars that were not presented in Paper~I. For that purpose, we adopt, as in Paper~I, a Sersic bulge
plus an exponential disk parametrization to describe the
two-dimensional surface brightness distribution of the lens galaxy
light profile. The Sersic profile is parametrized by means of the total
flux in the bulge, the Sersic index, $n$, the bulge ellipticity, $\epsilon$=1-b/a, the position angle of the bulge, PA,  and the effective
radius of the bulge, R$_{e}$.  The exponential profile depends on the
photometric disc total flux, the disc scale-length, the disc position
angle and the disc inclination. The software gives the best
fitting values for all of these parameters. The parameters of the Sersic
bulges are summarized in Table~\ref{priors}. The error bars correspond to the 68\% confidence level.
 For most systems, the results are consistent with the surface brightness parameters measured in Paper~I. But for others, such as COSMOS~5921+0638, the best fit parameters are different in Paper~I, Paper~III and here. Indeed, the presence of a ring or bright arcs close to the lens galaxy center makes difficult to produce a robust fit of the lens galaxy surface brightness density profile; this remains the case even when more complex fitting and deconvolution methods are used (see Chantry \& Magain 2007).

 The relative image positions to the lens galaxies are the main constraints for the lens models. For images which are point like objects, the determination of their positions only depends on the image resolution. This is the case for COSMOS~5921+0638. For this system the error on the relative position is  0.014\arcsec\, (see Paper~III).
 For the other multiple images systems, we determine the position of the 
the brightest peak in each image.  For theses systems, the error on the relative position of the images is typically 0.05\arcsec. For the perfect rings that do not display any peak we place the image arbitrarily around the ring, in a symmetric way around the lens center and we assume that the error on the  relative position of the images is 0.05\arcsec.  In
Table~\ref{tabexappen2} we provide the lens galaxy central coordinates as
well as the  position of the multiple images relative to the lens galaxy location used in the lens modeling.
For J100140.12+020040.9 and J095930.93+023427.7, we have retrieved the image position from Jackson 2009.

 \subsection{The mass models}\label{tmm}
 
 The purpose of the new lens modeling is: i) to measure the total mass of the galaxy within R$_E$: M($<$R$_E$) and ii) to check if the environment contribution measured in $\S$~\ref{lenscont} is a good estimator of the external shear .  

 To do  so, we have used the Lenstool code (Kneib et al. 1993, Jullo et al. 2007)
to model the lens galaxy mass distributions.  Lenstool allows a
$\chi^2$ minimization of parametric mass models, either in the source
or in the image plane. For higher accuracy, we use the image plane
minimization algorithm.  

\begin{table*}[ht]
\centering
\caption{\label{SIE} Best fit parameters for the lens models: SIE + shear. }\begin{tabular}{ c c r c  c  r  r r r  }
\hline
\hline
Name   & N&$\chi^2$  & ($\chi^{2\prime}$, $\gamma$, PA$_\gamma$) &$\sigma_v\pm\delta\sigma_v$& R$_E$ & R$_E$& M($<$R$_E$)& f$_{DM}$($<$R$_E$)\\
                 & &             &   &     {\it km~s$^{-1}$} & {\it \arcsec} &{\it kpc} & {\it 10$^{11}$M$_\odot$}& \\
\hline
COSMOS~0049+5128&R &35.3    & (5.6, 0.023, +83$\degree)$ & 313$^{+7}_{-10}$ &   2.17$^{+0.00}_{-0.06}$&10.2$^{+0.0}_{-0.3}$ & 7.31$^{+0.11}_{-0.60}$ & 0.94$^{+0.06}_{-0.03}$\\
COSMOS~5947+4752&R & 1.6    & \_ &326$^{+11}_{-6}$ & 2.33$^{+0.00}_{-0.00}$ &11.1$^{+0.0}_{-0.0}$  &8.61$^{+0.59}_{-0.32}$ & 0.95$^{+0.06}_{-0.03}$\\
COSMOS~5921+0638&4& 11.8  & (2.0, 0.025, +20.5)& 189$^{+0}_{-1}$ &0.62$^{+0.00}_{-0.00}$ &4.4$^{+0.0}_{-0.0}$ &1.10$^{+0.01}_{-0.00}$&0.58$^{+0.00}_{-0.0-}$ \\
COSMOS~0038+4133&4 &6.3   &  (0.5, 0.128, -76.3$\degree$) & 207$^{+23}_{-12}$&  0.60$^{+0.00}_{-0.00}$&4.3$^{+0.0}_{-0.0}$ &1.41$^{+0.31}_{-0.21}$&-0.37$^{+0.05}_{-0.07}$\\
COSMOS~0124+5121&R & 0.9  &\_  &267$^{+38}_{-19}$& 0.91$^{+0.00}_{-0.01}$ & 6.8$^{+0.0}_{-0.1}$ &3.55$^{+1.0}_{-0.52}$ &0.84$^{+0.24}_{-0.12}$ \\
COSMOS~0047+5023& 4& 1283.1 &(10.4, 0.126, -15.10$\degree$) & 383$^{+9}_{-38}$&1.80$^{+0.00}_{-0.39}$ &13.5$^{+0.0}_{-2.9}$&14.52$^{+0.02}_{-3.63}$ &0.68$^{+0.00}_{-0.17}$  \\
J100140.12+020040.9&4 &7.3 &(2.3, 0.053, +0.16$\degree$) &259$^{+43}_{-18}$ & 0.81$^{+0.01}_{-0.00}$&6.1$^{+0.1}_{-0.0}$ &3.00$^{+1.10}_{-0.41}$ & 0.60$^{+0.22}_{-0.08}$\\
COSMOS~5941+3628&R &0.5 & \_ &315$^{+51}_{-20}$ & 1.23$^{+0.02}_{-0.00}$& 9.3$^{+0.1}_{-0.0}$&6.80$^{+2.37}_{-0.80}$&0.62$^{+0.24}_{-0.07}$\\
J095930.93+023427.7&4 &6.8& \_& 255$^{+37}_{-17}$& 0.79$^{+0.00}_{-0.02}$&6.0$^{+0.0}_{-0.2}$ &  2.89$^{+0.77}_{-0.38}$  &0.71$^{+0.30}_{-0.12}$\\
COSMOS~0050+4901&4&545.4& (26.3, 0.097, +79.9$\degree$)  &386$^{+82}_{-30}$& 1.65$^{+0.02}_{-0.00}$&12.7$^{+0.2}_{-0.0}$&11.88$^{+8.99}_{-0.06}$&0.92$^{+0.55}_{-0.18}$ \\
COSMOS~0018+3845&3  &45.3& (0.2, 0.103, +39.9$\degree$)& 289$^{+2}_{-0}$  & 1.32$^{+0.01}_{-0.00}$&10.2$^{+0.1}_{-0.0}$ &6.22$^{+0.15}_{-0.00}$ &0.93$^{+0.00}_{-0.01}$\\
COSMOS~5914+1219&3&215.9& (0.1, 0.086, +71.2$\degree$) & 358$^{+23}_{-14}$& 1.60$^{+0.00}_{-0.01}$&12.8$^{+0.0}_{-0.1}$ & 12.00$^{+1.51}_{-0.87}$ &0.79$^{+0.13}_{-0.04}$\\
\hline
\end{tabular}
\tablefoot{(1)  Lens name. (2) Number of images or ``R'' if it is a complete Einstein ring. (3) $\chi^2$ of the best lens modeled obtained when using the priors on the external shear measured in $\S$~\ref{lenscont}. (4) Favorite shear parameters (when the one measured in $\S$~\ref{lenscont} lead to $\chi^{2}>>$1) and corresponding $\chi^{2}$. (5) Velocity dispersion of the SIE corresponding to the best model in (4) or in (3) if no other.  (6) Corresponding Einstein radius in arc-second and (7) in kpc.  (8) Mass of the lens galaxy in the Einstein radius. (9) Fraction of DM in the Einstein radius. In column (5), (6), (7) and (8), the errors are due to the source redshift uncertainty (as explained in $\S$~\ref{lenscont}). }

\end{table*}

\subsubsection{SIE+$\gamma$} \label{siegamma}
 We have performed a set of lens model in which the mass distribution the lens galaxies follows that of a Singular Isothermal Ellipsoid (SIE).  A SIE is defined by its position, velocity dispersion, orientation and ellipticity.  Assuming that mass follows light, we fix the SIE central position to that of the galaxy light profile (Table~\ref{tabexappen2}). We chose its  orientation to be that of the Sersic bulge light profile  (Table~\ref{priors}) with an additional $\pm$10$\degree$ uncertainty to take into account for a possible misalignment between the luminous bulge and the DM halo (e.g. Kochaneck 2002).
 The  higher boundary for the SIE ellipticity   is set equal to that of the Sersic profile (Table~\ref{priors}); the lowest boundary is set to 0 in order to take into account for a possibly shallower distribution of the DM halo in comparison to the luminous core  observed in ETGs (e.g. Gavazzi et al. 2007).
 
  In addition,  we model the contribution of the environment  by an ``external shear'' which is parametrized by a shear strength and orientation. The group and cluster contributions to the shear at the lens galaxy locations are measured in $\S$~\ref{lenscont}. We use the values of $\gamma$ and PA$_\gamma$  from Table~\ref{knobfig}, and the corresponding error bars as priors, to build the lens models.
  
\begin{figure*}[ht]
\begin{center}
\includegraphics[width=7cm]{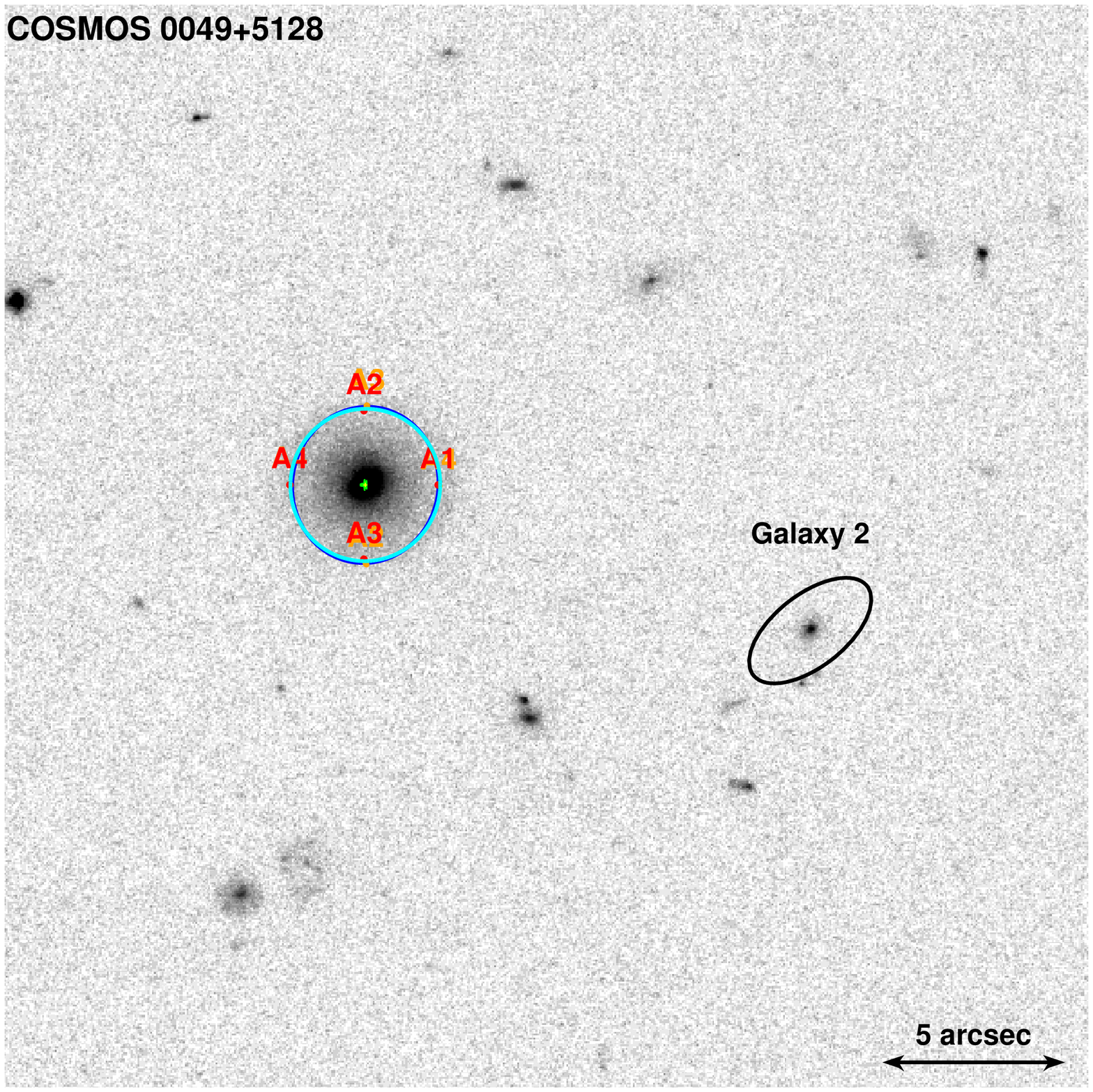}\includegraphics[width=7cm]{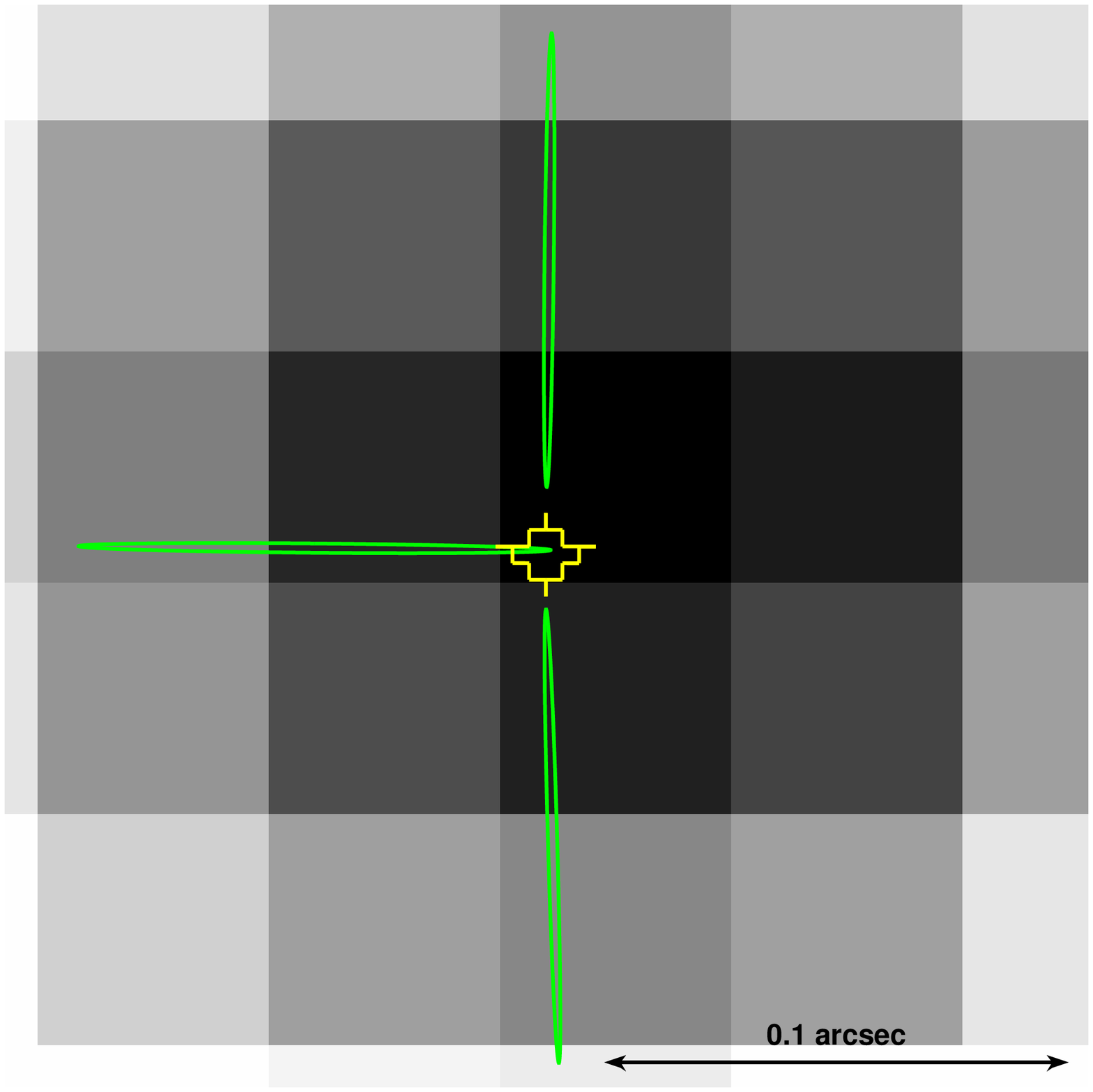}
\includegraphics[width=7cm]{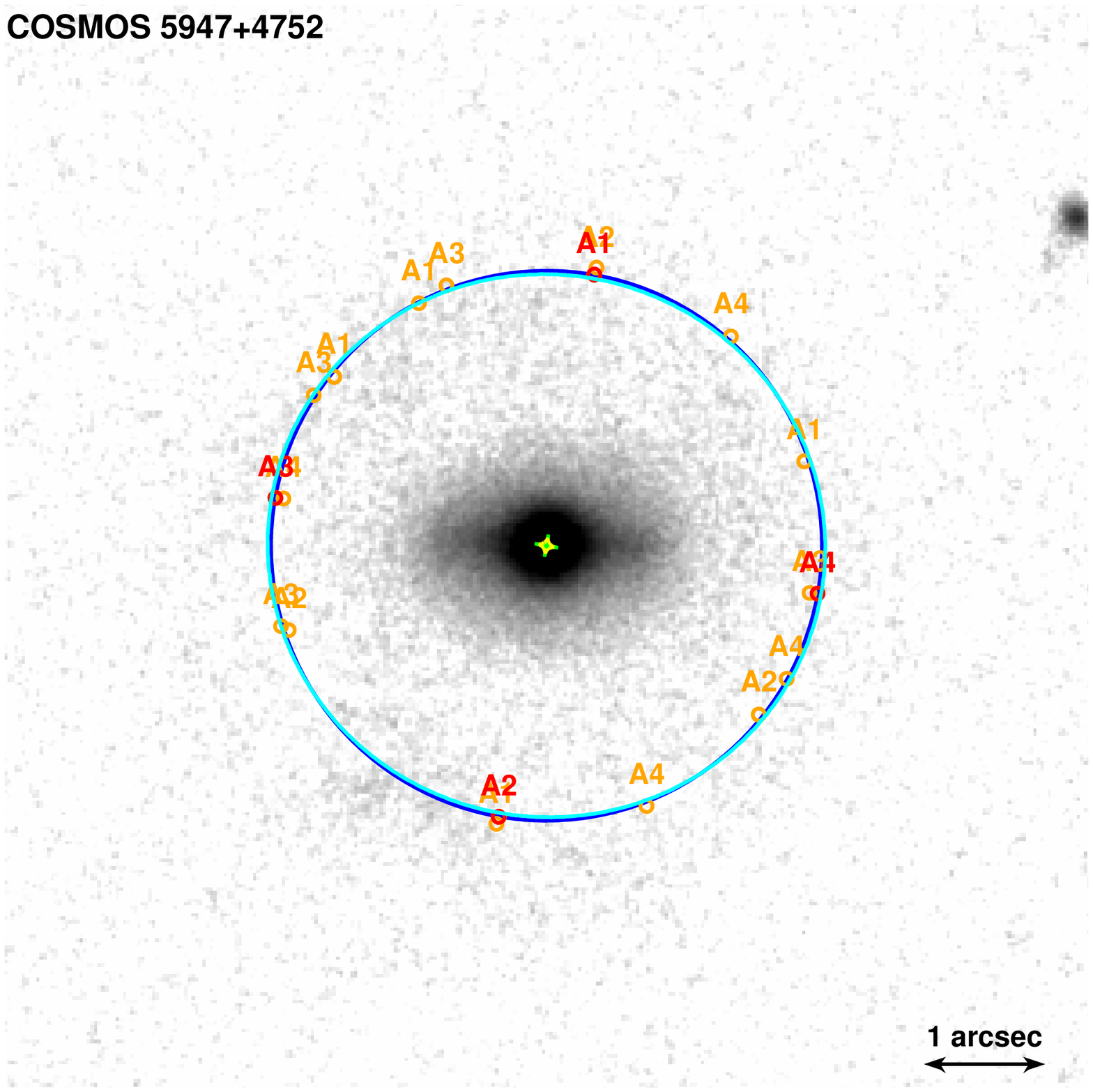}\includegraphics[width=7cm]{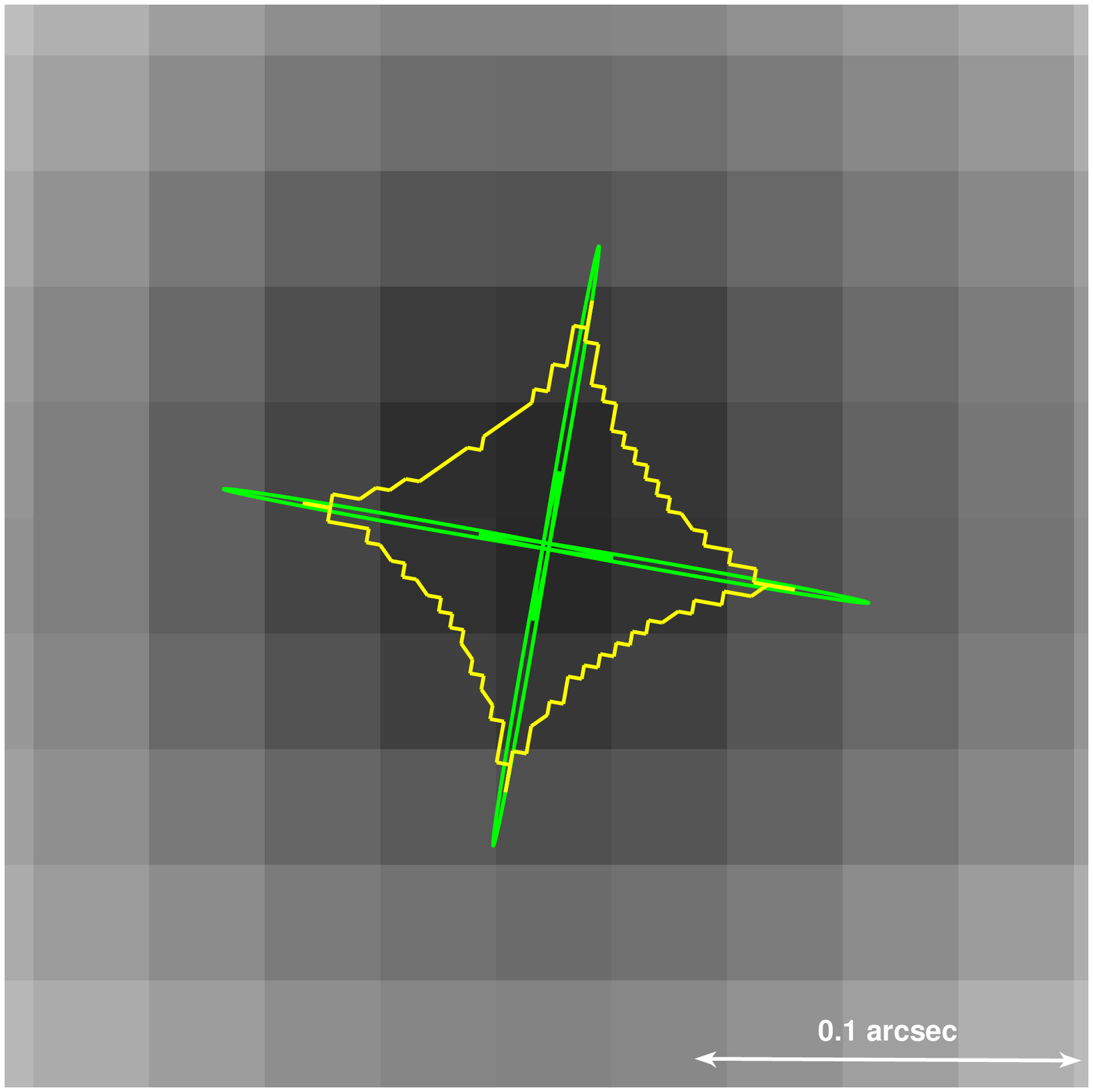}
\caption{Mass models on top of ACS images for (top to bottom) COSMOS~0049+5128 and COSMOS~5947+4752. The left column displays the  best mass model corresponding to the $\chi^2$ given in Table~\ref{SIE} (column 4 or column 3 if no other).  The right panel shows a zoom of the central region of the lens models, where the source can be seen. North is to the top and East to the left. 
  Color code: the red circles are the observed images, which radii correspond to the position uncertainty used in the modeling. In orange are the images
  produced by the best lens model (in case of perfect fit, the orange
  and red crosses superimpose).  The navy blue lines describe the
  potential. The caustic lines are in yellow an
  the critical lines are in cyan. The green ellipses show the position of
  the source as seen through the best mass model (one source for each image, when a good fit is reached
  the four sources are partially superimposed). We indicate the position of Galaxy 2 (see  $\S$~\ref{siegamma} and Table~\ref{shear}).}
\label{mmodel}
\end{center}
\end{figure*}

\begin{figure*}[ht]
\begin{center}
\includegraphics[width=7cm]{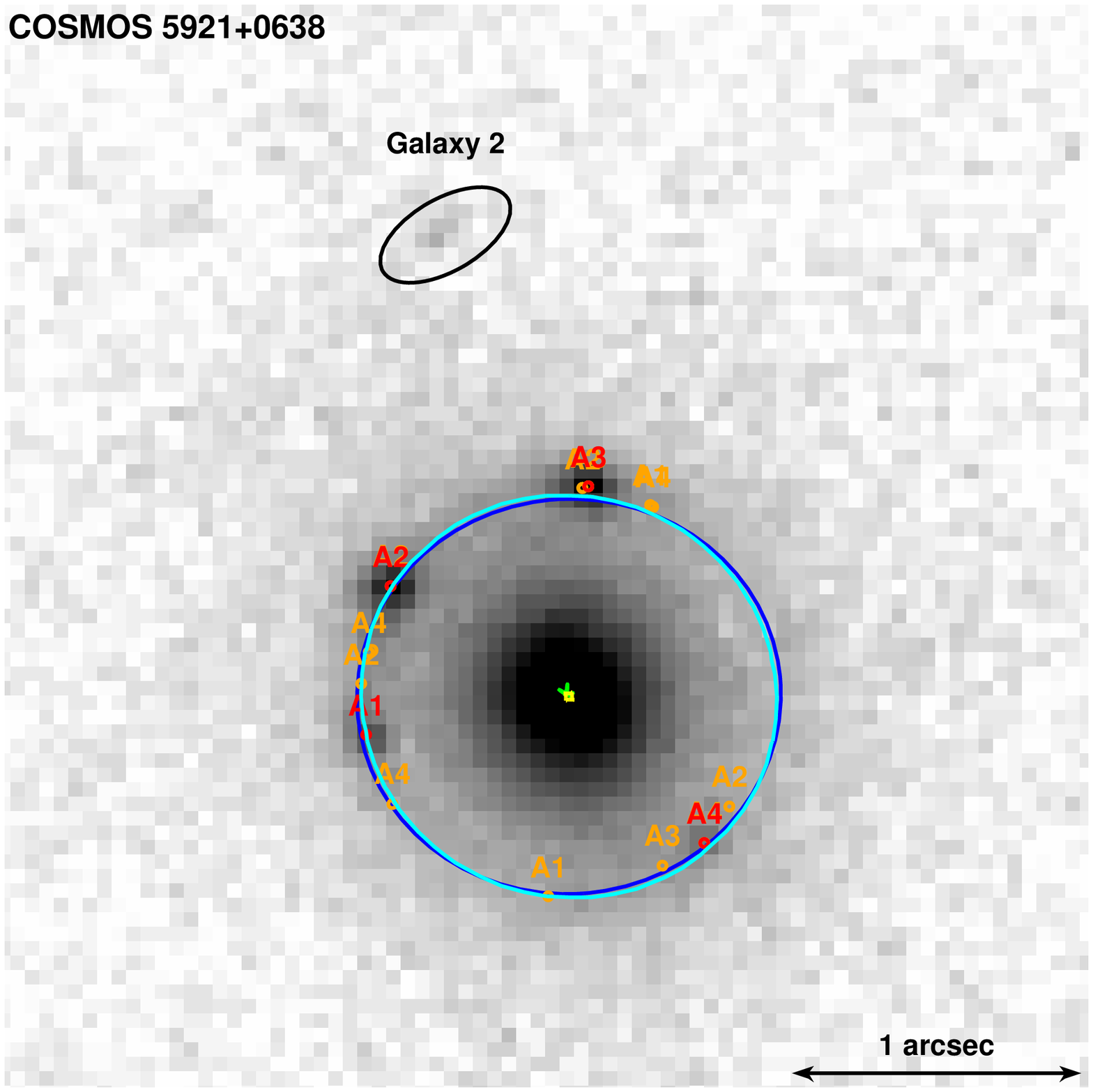}\includegraphics[width=7cm]{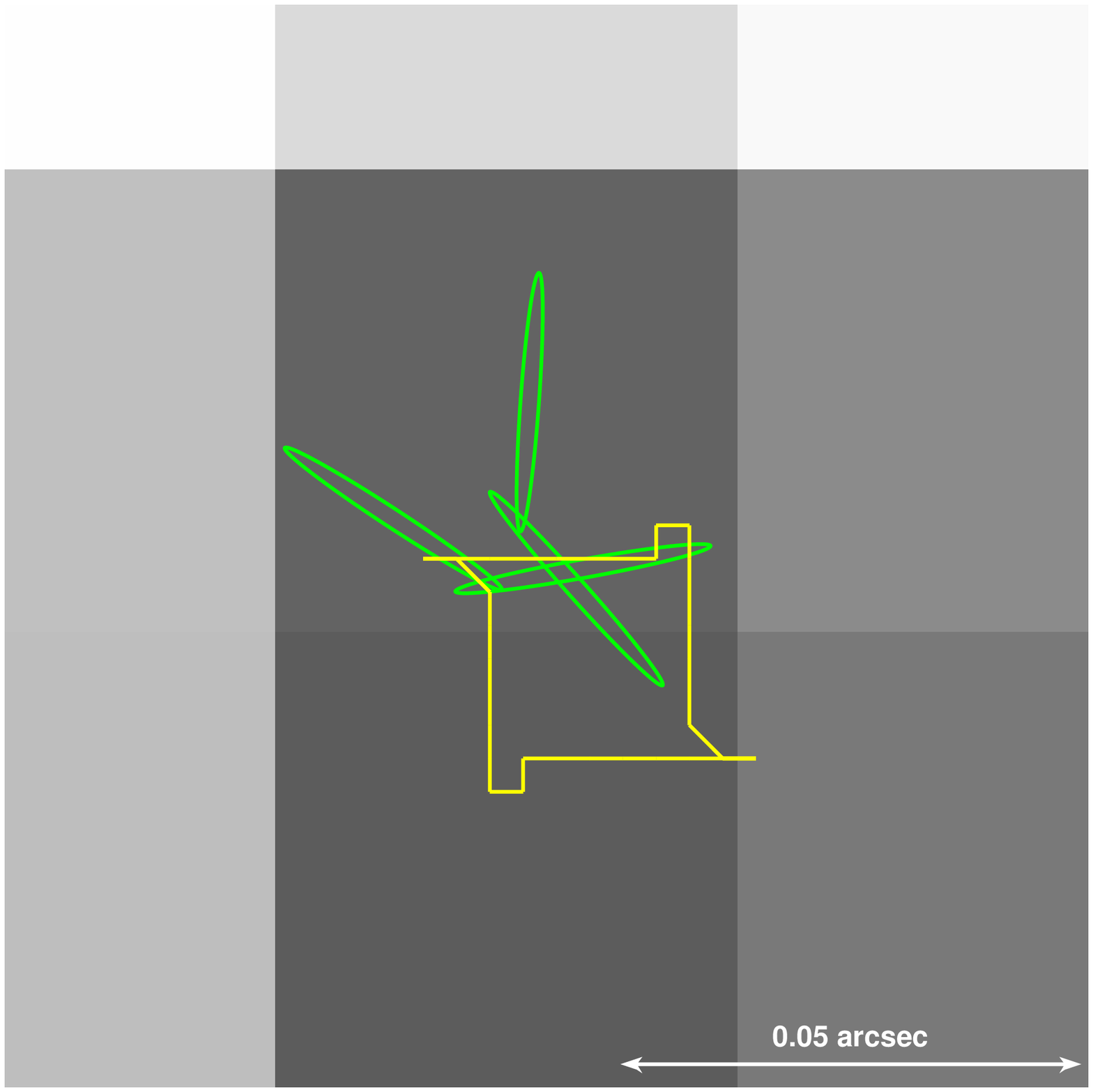}
\includegraphics[width=7cm]{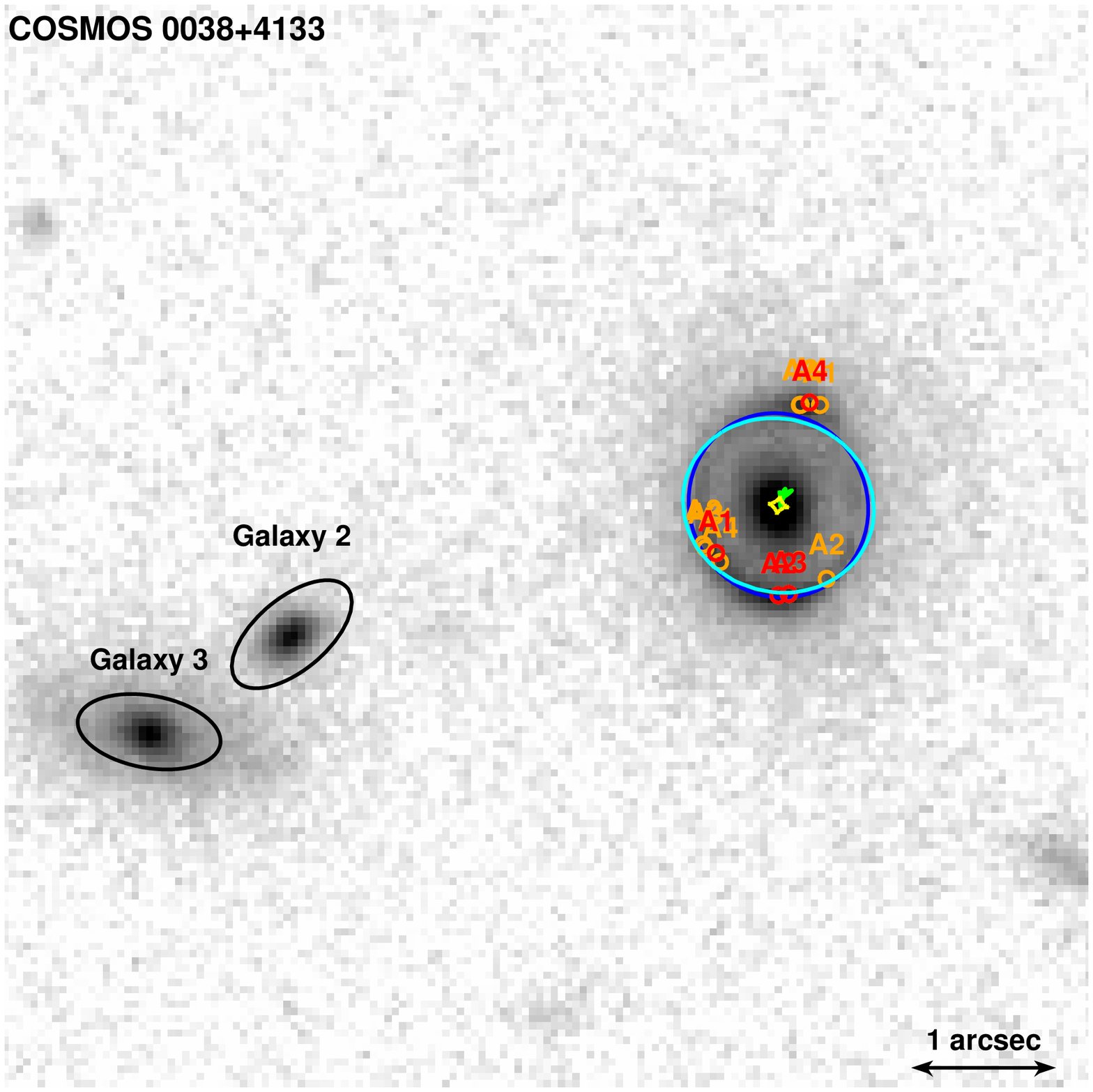}\includegraphics[width=7cm]{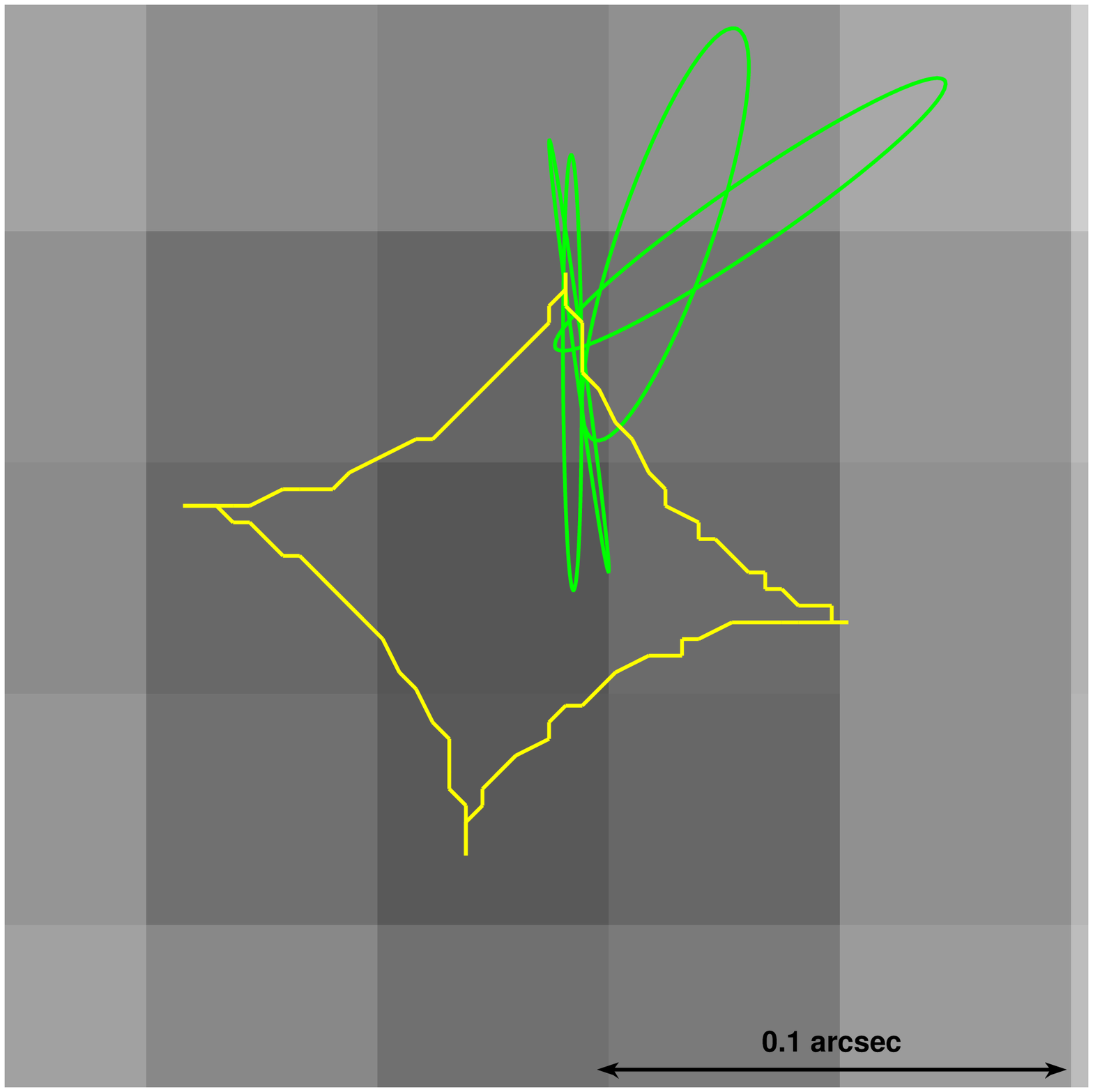}
\includegraphics[width=7cm]{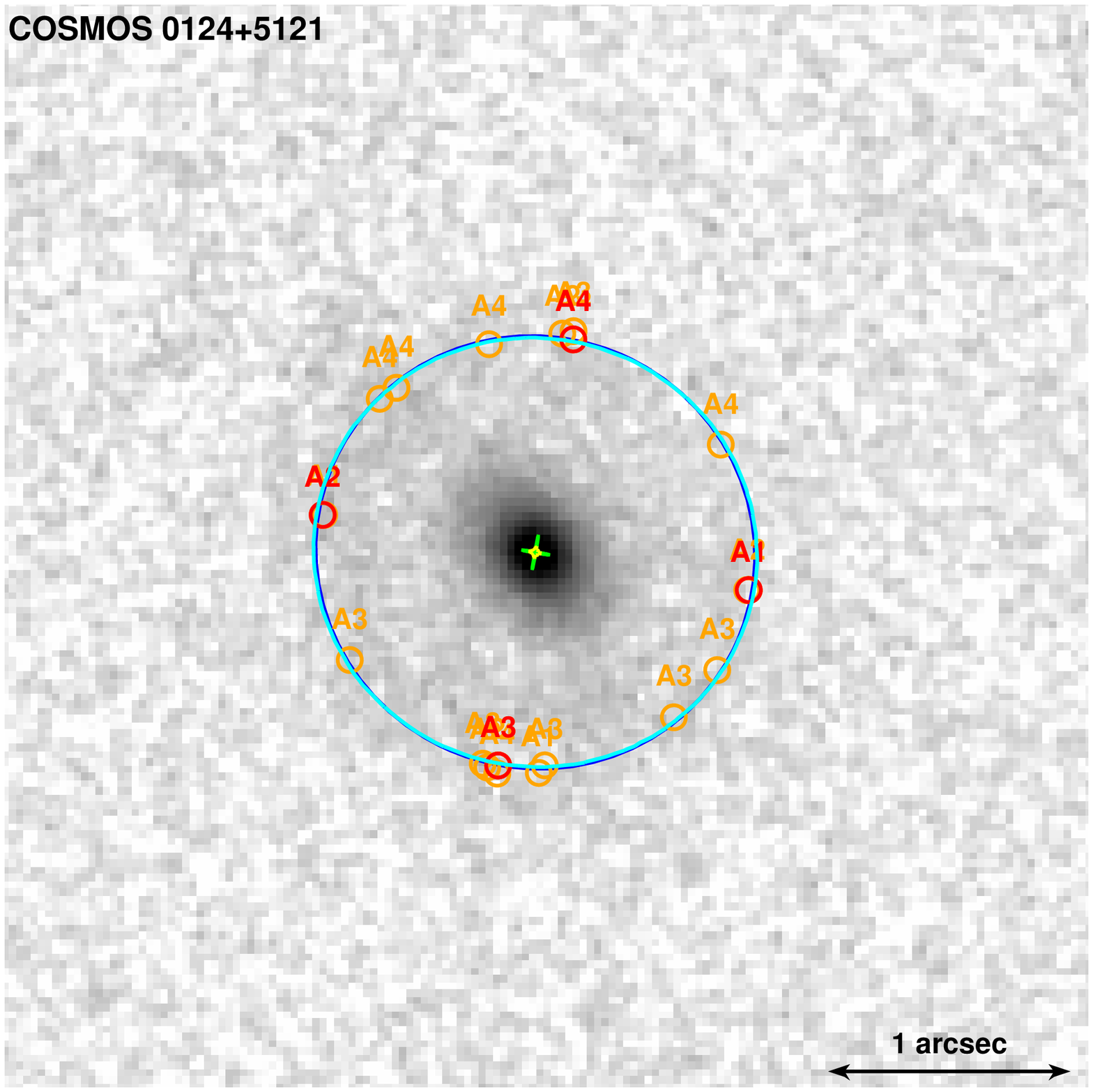}\includegraphics[width=7cm]{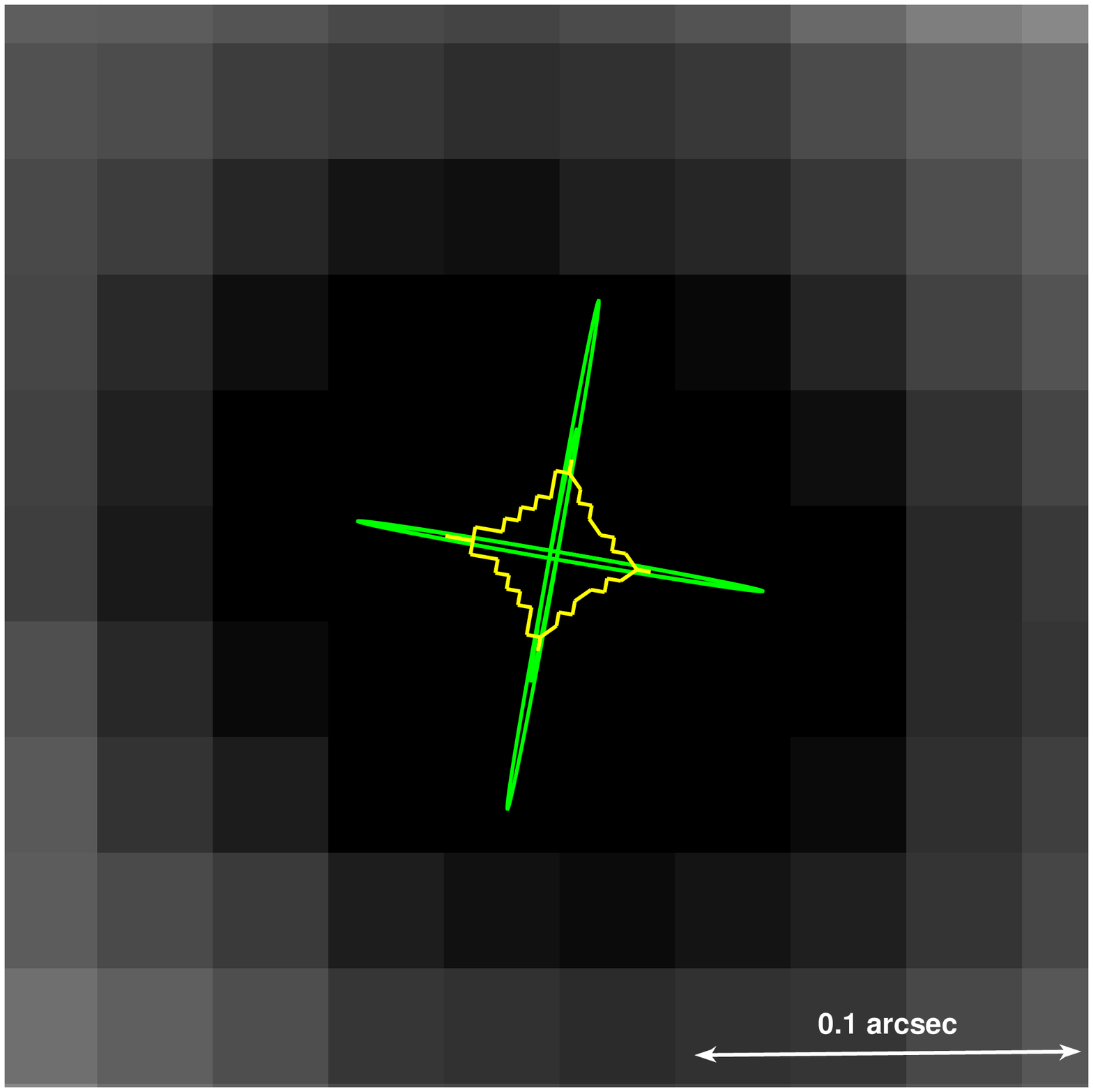}
\caption{ Suite: Mass models on top of ACS images for (top to bottom)  COSMOS~5921+0638, COSMOS~0038+4133 and COSMOS~0124+5121.  The color code for the labels is given in Fig.~\ref{mmodel}. North is to the top and East to the left. }
\label{mmodel2}
\end{center}
\end{figure*}

\begin{figure*}[ht]
\begin{center}
\includegraphics[width=7cm]{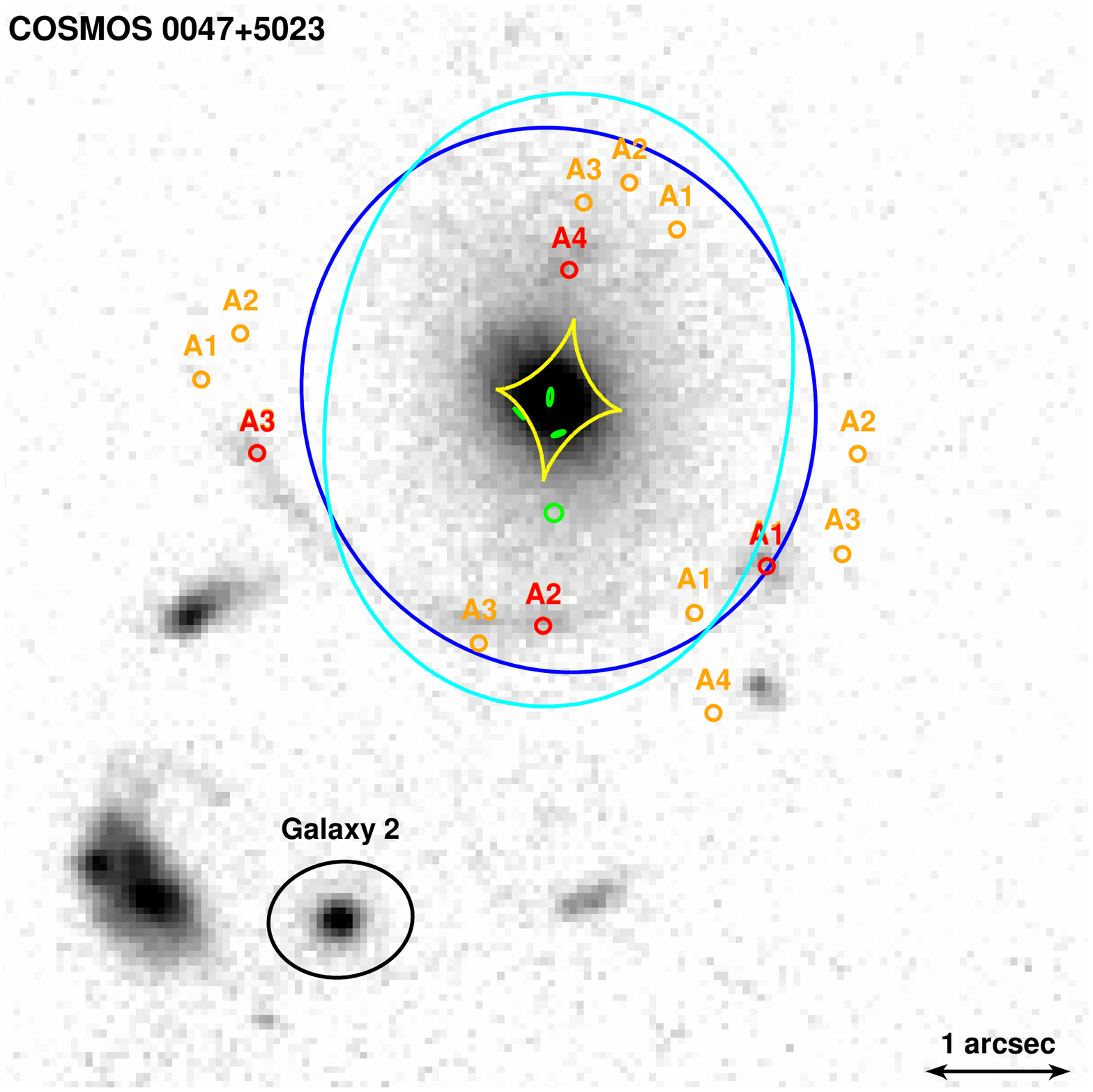}\includegraphics[width=7cm]{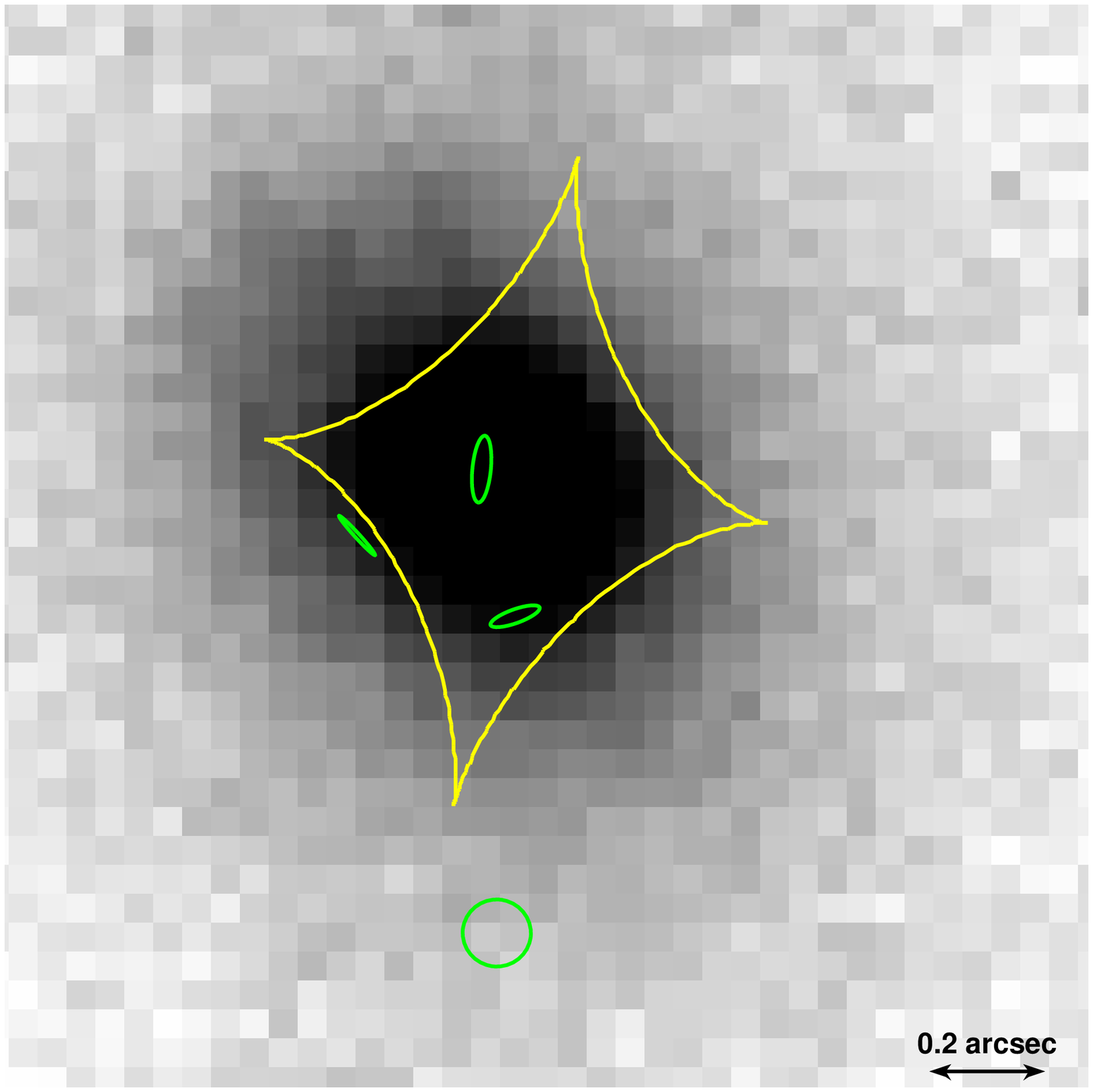}
\includegraphics[width=7cm]{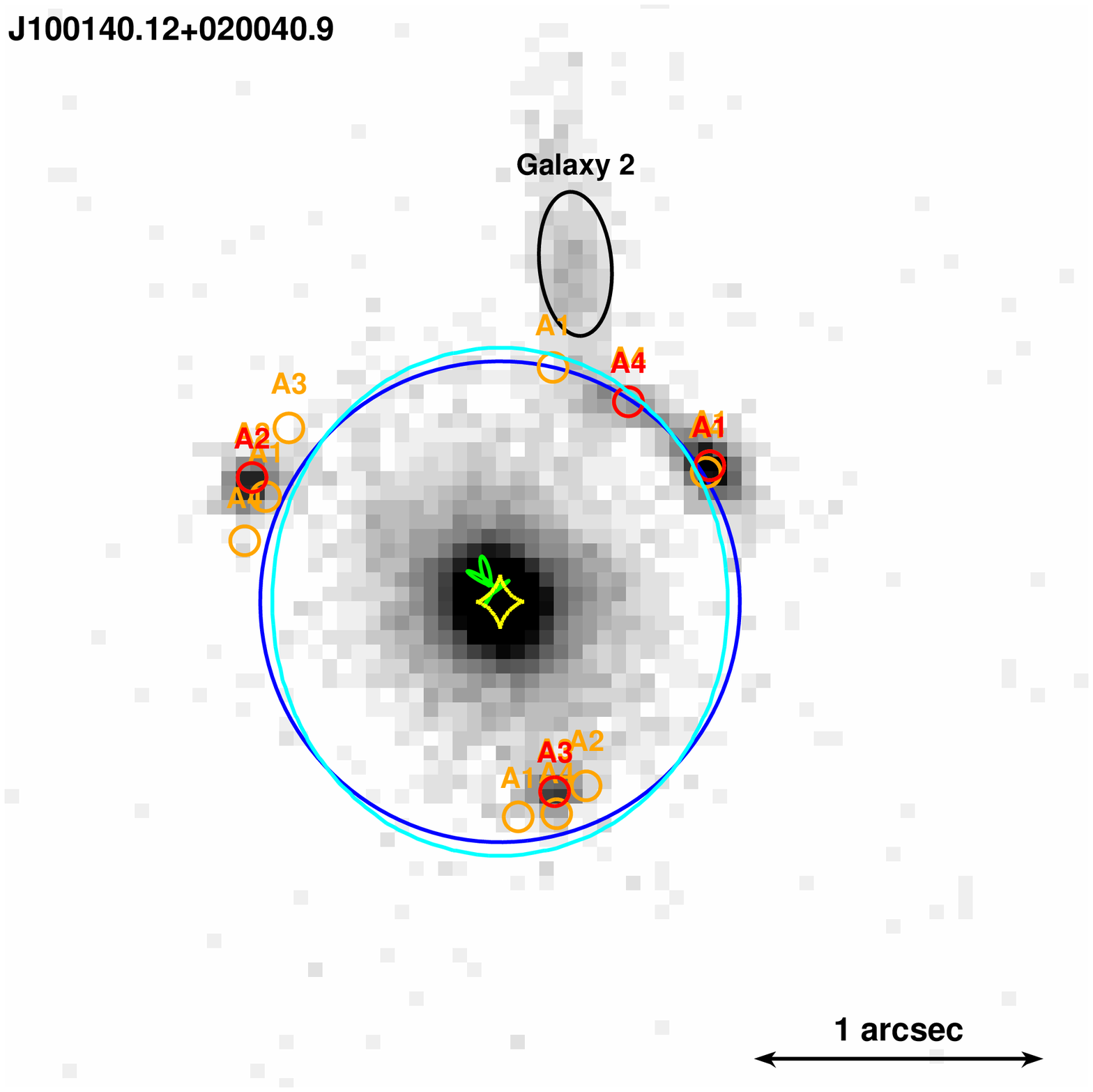}\includegraphics[width=7cm]{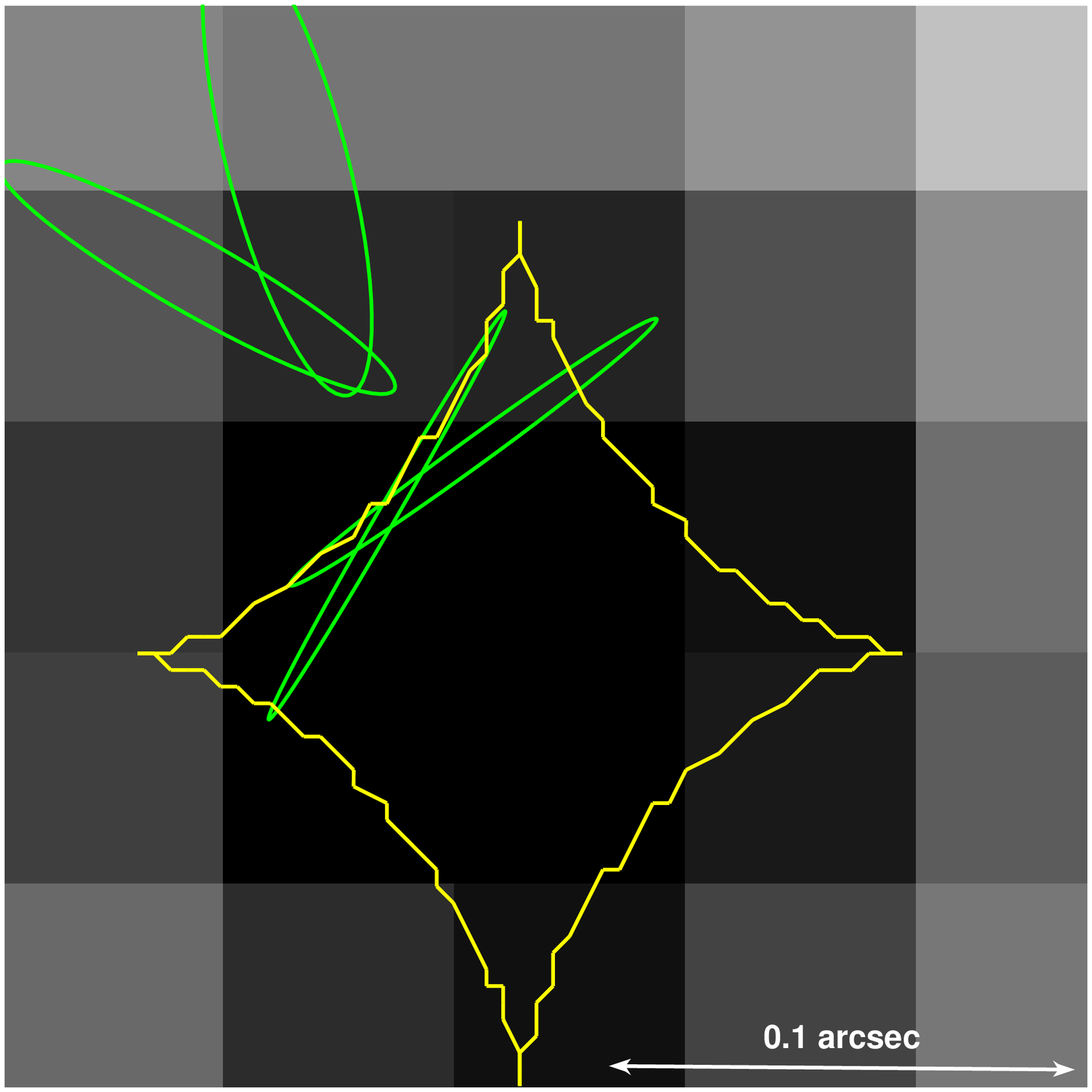}
\includegraphics[width=7cm]{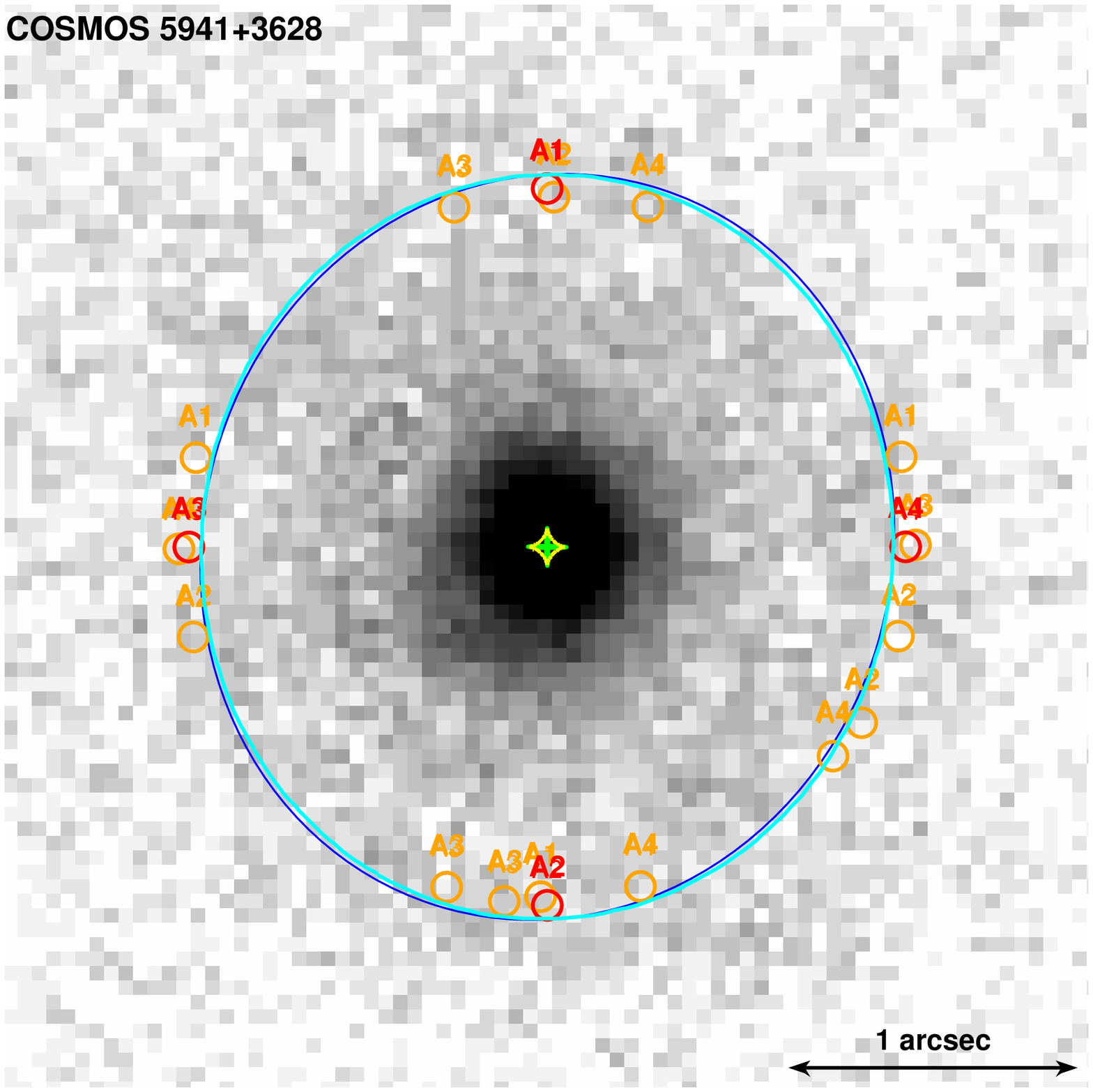}\includegraphics[width=7cm]{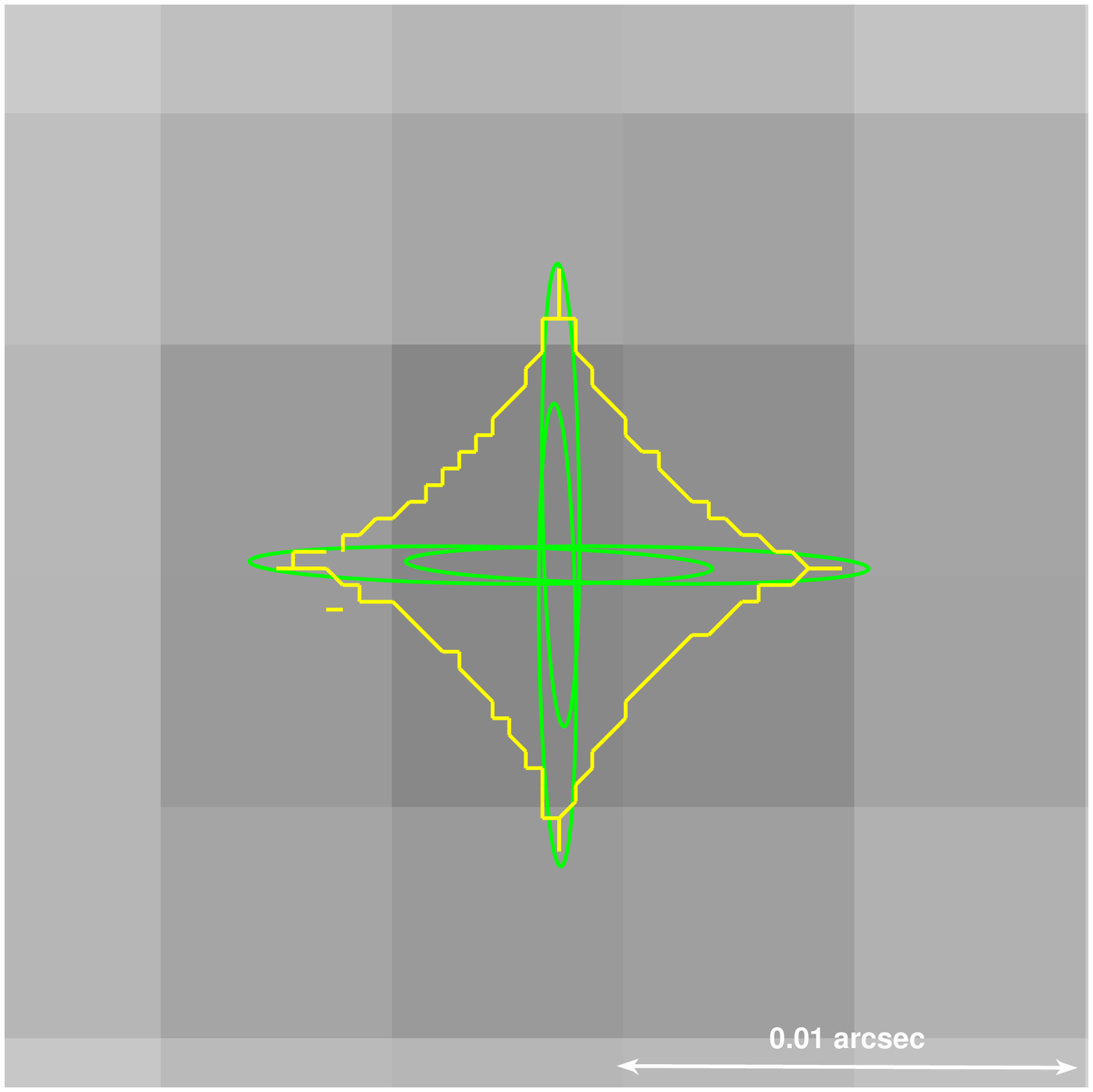}
\caption{ Suite: Mass models on top of ACS images for (top to bottom) COSMOS~0047+5023, J100140.12+023427.7 and COSMOS~5941+3628.  The color code for the labels is given in Fig.~\ref{mmodel}. North is to the top and East to the left. }
\label{mmodel3}
\end{center}
\end{figure*}

\begin{figure*}[ht]
\begin{center}
\includegraphics[width=7cm]{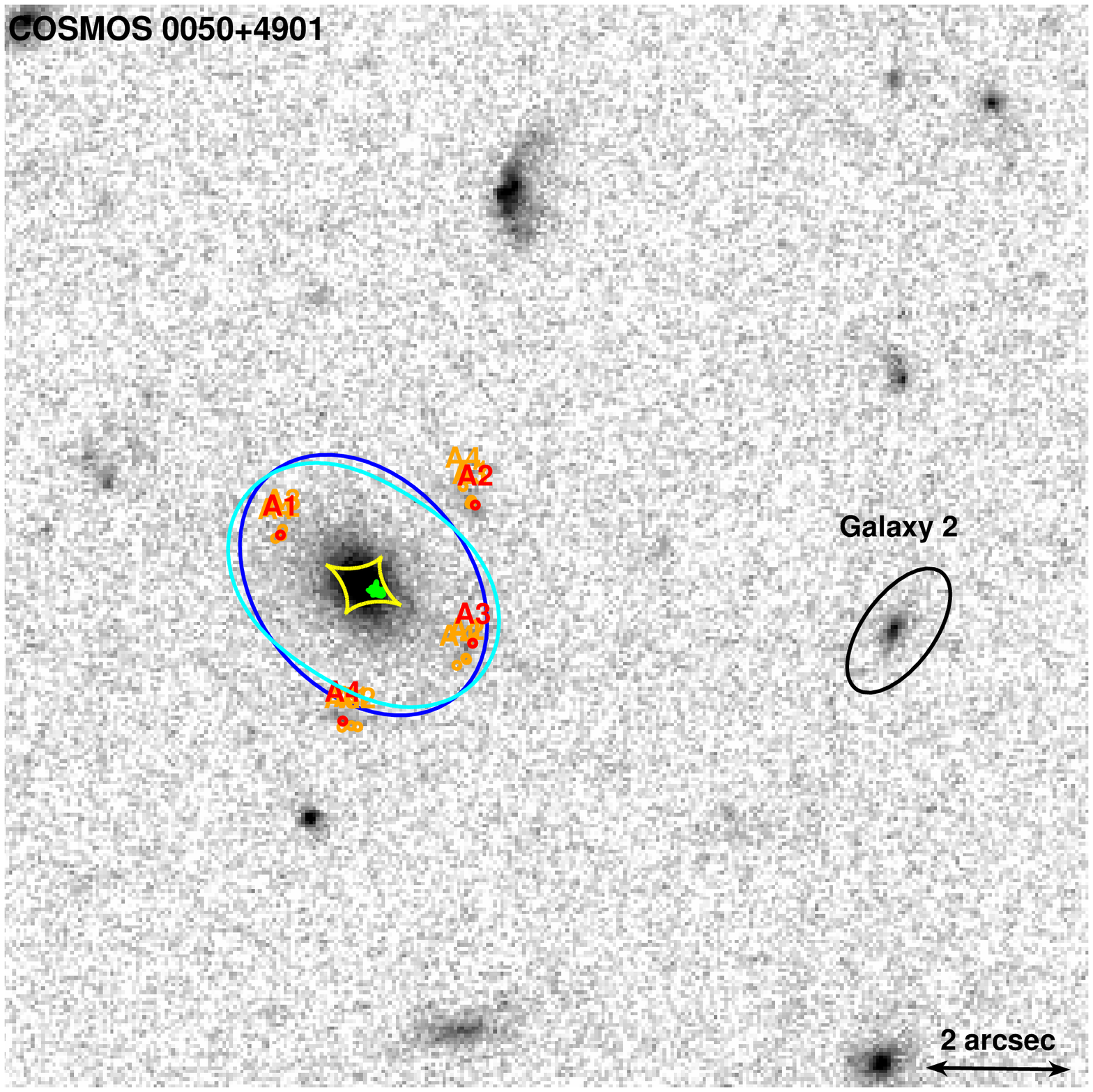}\includegraphics[width=7cm]{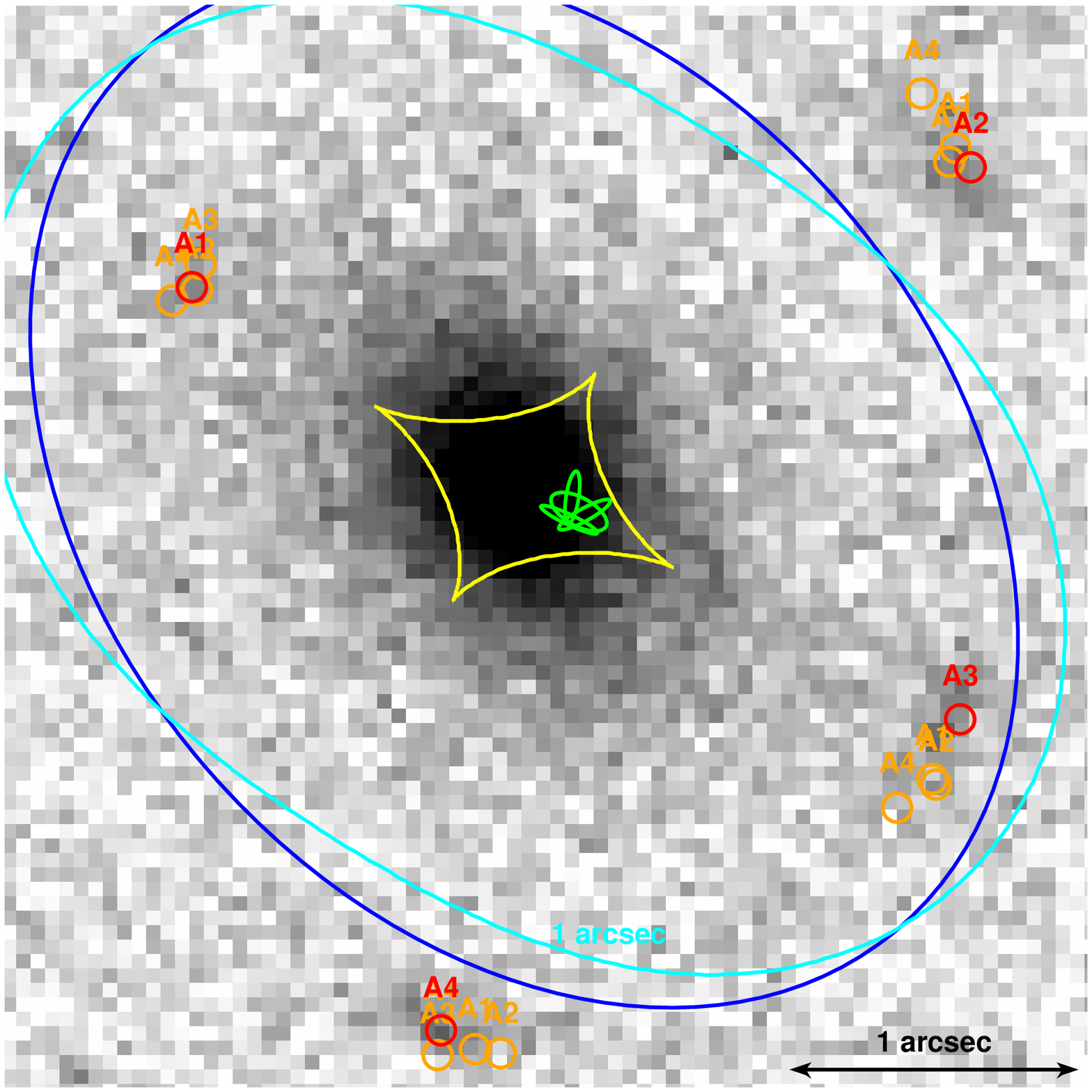}
\includegraphics[width=7cm]{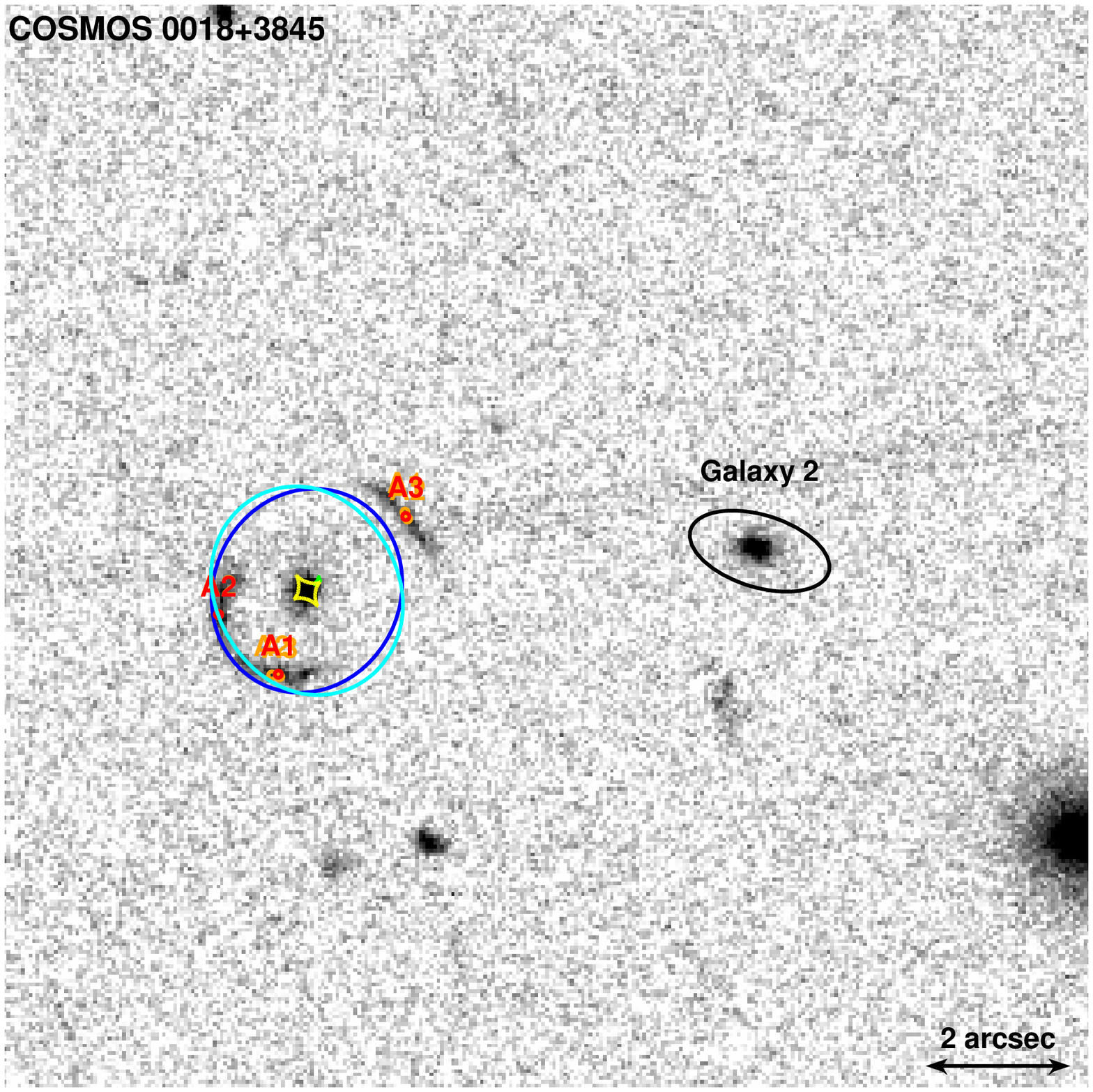}\includegraphics[width=7cm]{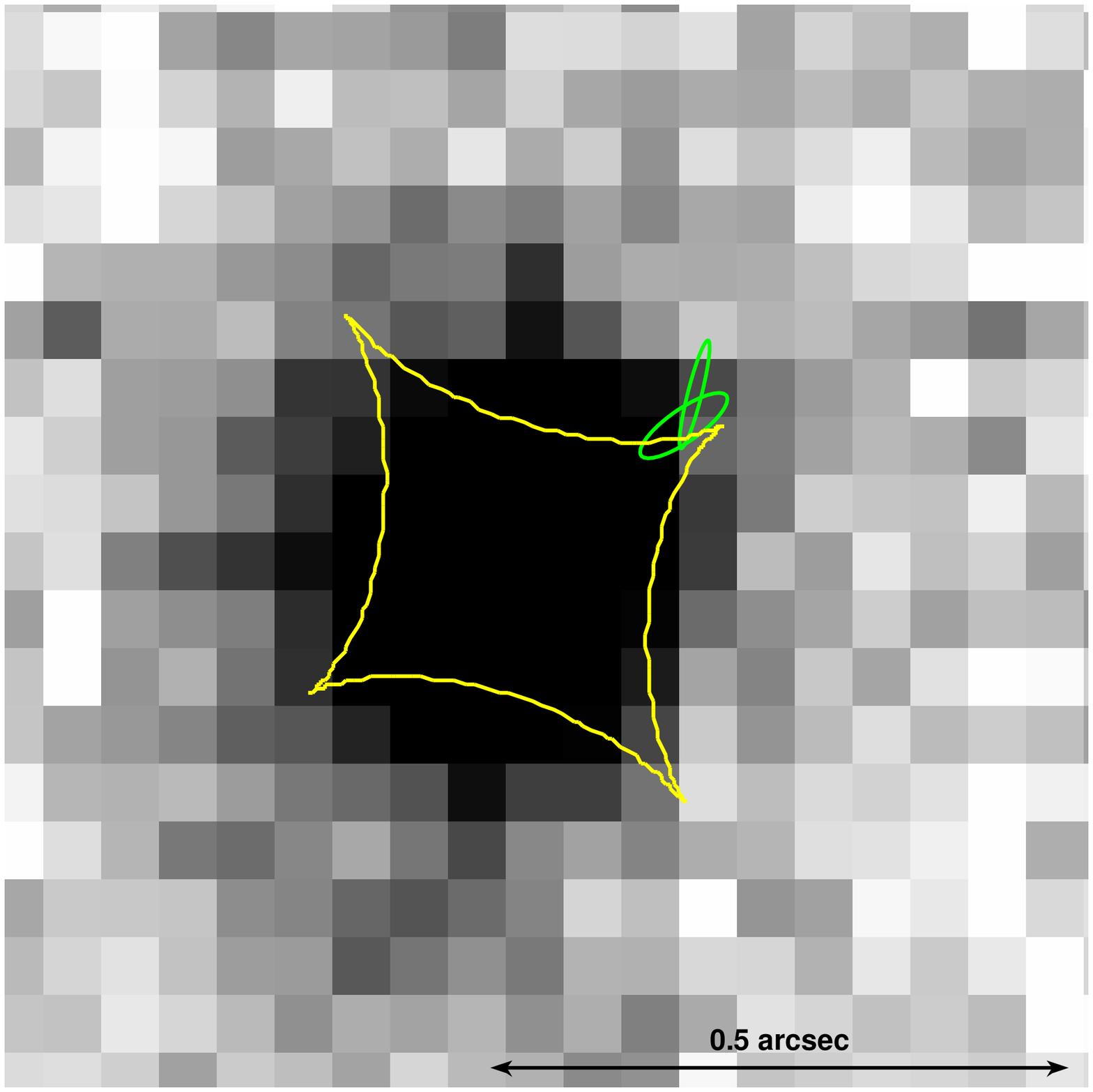}
\includegraphics[width=7cm]{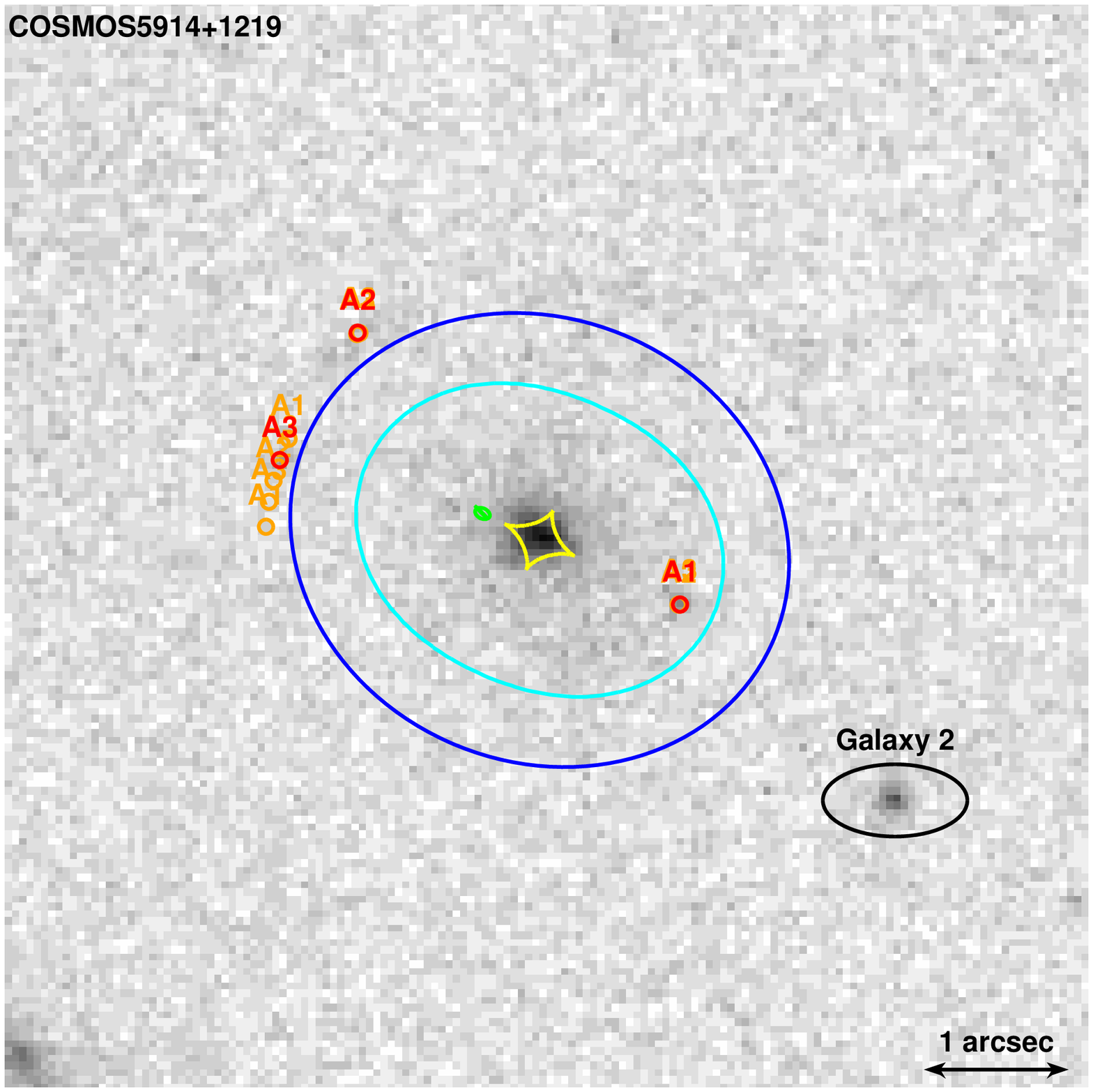}\includegraphics[width=7cm]{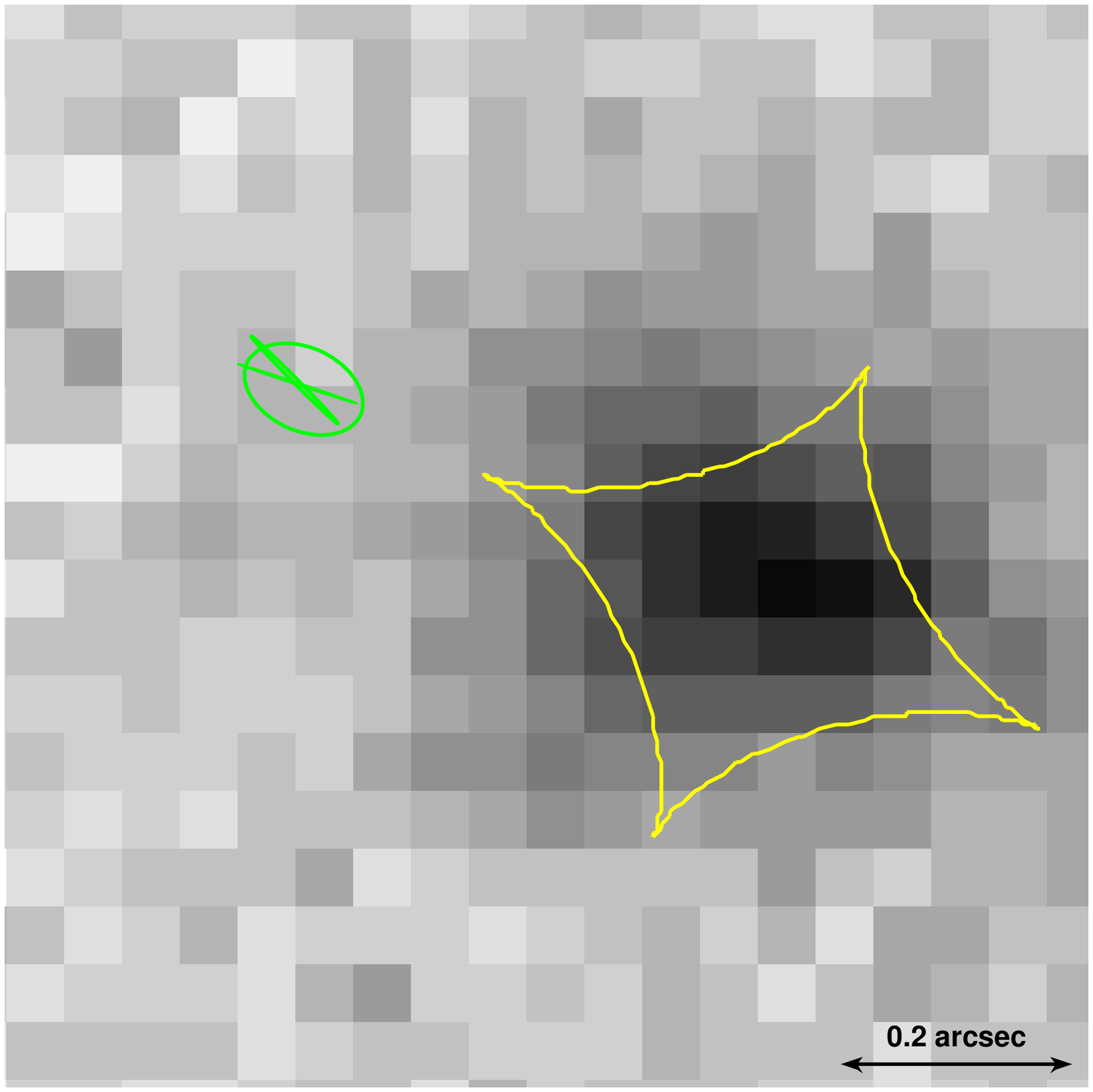}
\caption{ Suite: Mass models on top of ACS images for (top to bottom) COSMOS~0050+490, COSMOS~0018+3845 and  COSMOS~5914+1219. The color code for the labels is given in Fig.~\ref{mmodel}. North is to the top and East to the left. }
\label{mmodel4}
\end{center}
\end{figure*}

\begin{figure*}[ht]
\begin{center}
\includegraphics[width=8cm]{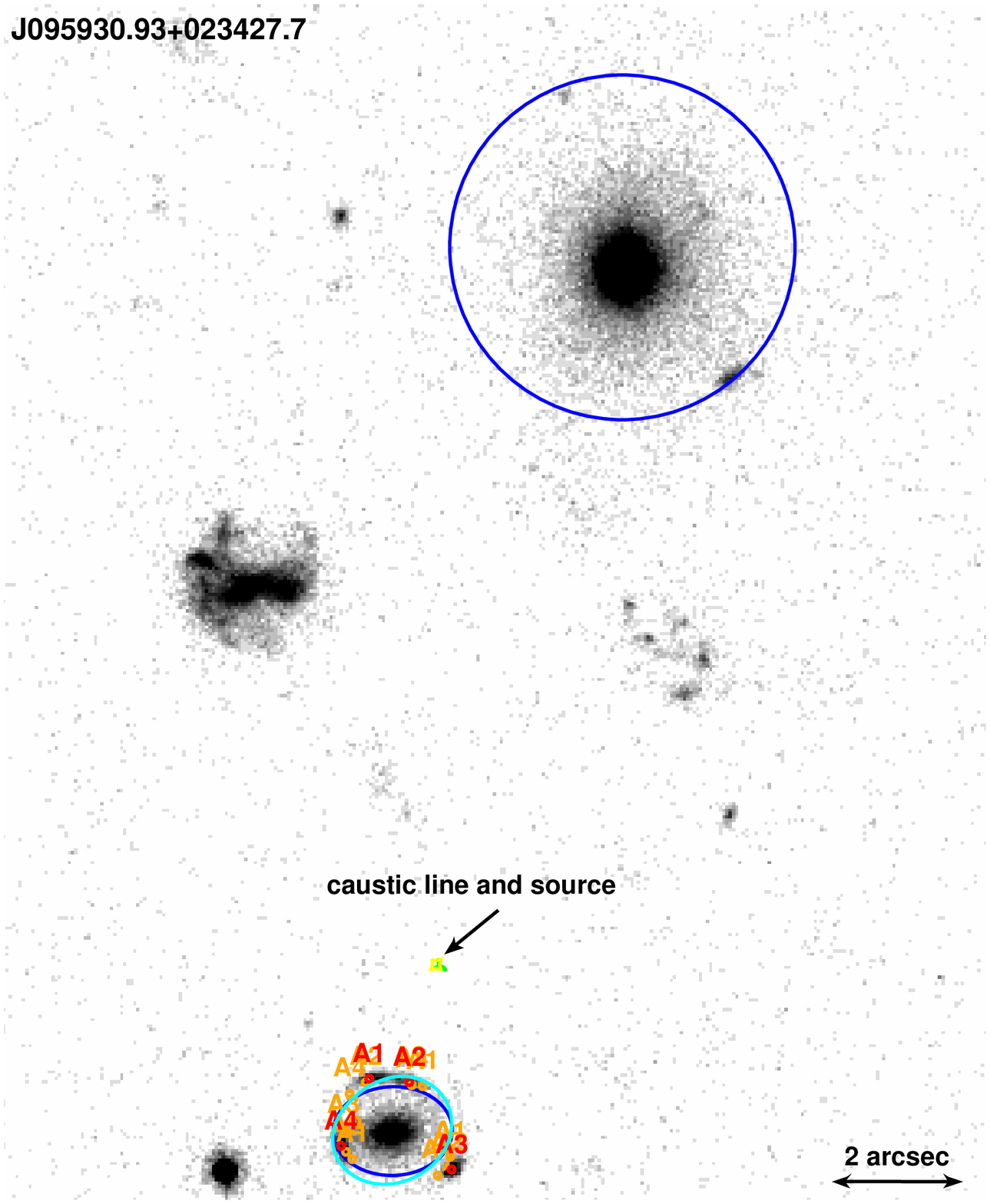}\includegraphics[width=7cm]{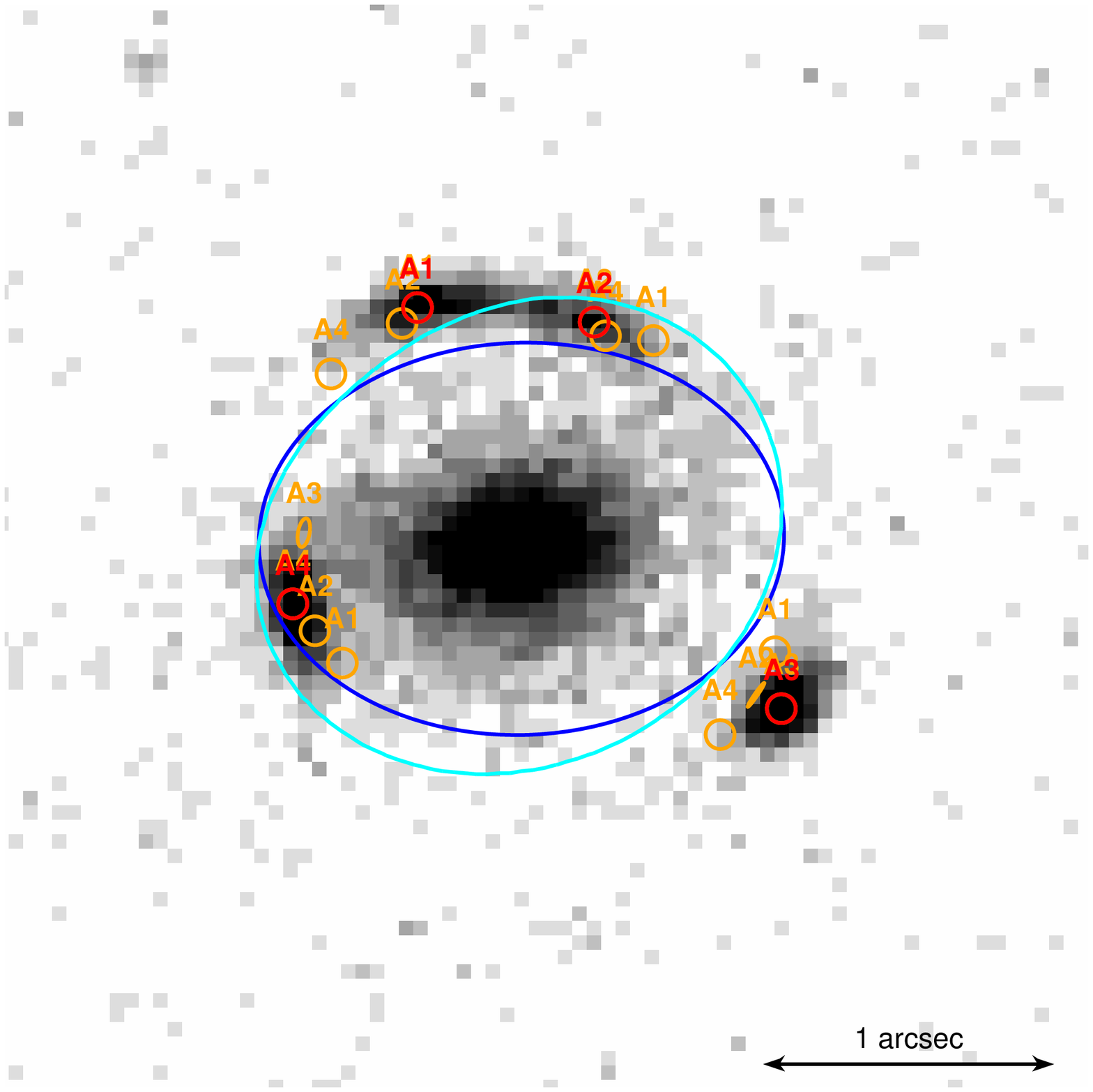}
\includegraphics[width=7cm]{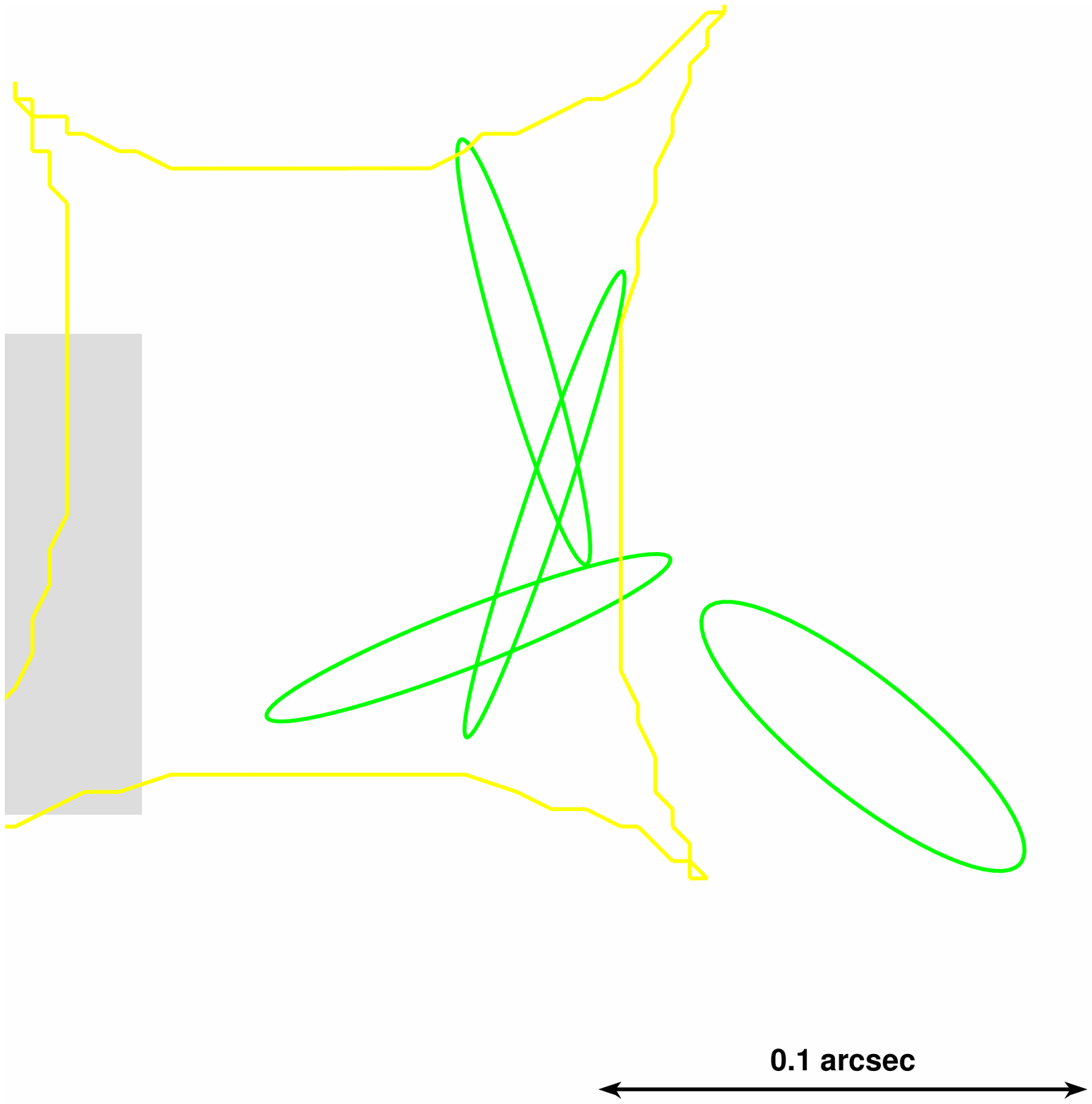}
\caption{ Suite: Mass models for the J095930.93+023427.7.  Top left panel: the navy blue circle located North of the image surrounds the group central galaxy.  Top right panel: zoom on the lens galaxy and on the images. Bottom panel: zoom on the caustic (yellow curves) and on the favorite source position (green ellipses). The color code for the labels is given in Fig.~\ref{mmodel}. North is to the top and East to the left.  }
\label{mmodel5}
\end{center}
\end{figure*}

     Therefore, the  total number of free parameters for the models is 5: 3 for the SIE (orientation, ellipticity, velocity dispersion) and 2 for the external shear (strength and direction). The number of observational constraints is 6 for the quads and rings and 4 for the triple image lenses. 
 The $\chi^2$ of the best  lens models are reported in Table~\ref{SIE} (column~3). The best fit models (with $\chi^2$ in column~3) are displayed in Figs.~\ref{mmodel} to \ref{mmodel5}. Whenever $\chi^2 >> 1$, we have performed a second lens model, changing the priors on the external shear to the following ones:   $\gamma$=[0.0,0.9], PA$_\gamma$=[-90$\degree$,90$\degree$], hence letting the external shear parameters free. The $\chi^2$ of these second lens models are reported in Table~\ref{SIE} (column~4) and referred as $\chi^{2\prime}$.  
 
 For J095930.93+023427.7, we have modeled the group in direction to the lens by a SIS which position is fixed to the position of the group in the X-ray catalog. Hence, the only parameter allowed to vary is the velocity dispersion of the profile. The best fit model in Table~\ref{SIE} is obtained for a group with velocity dispersion: $\sigma_v^{group}$=408$^{+94}_{-29}$~km~s$^{-1}$, if it was at the lens galaxy redshift.  From  Fig.~\ref{mmodel5}, we see that the fit is not perfect: with additional observational constraints such as the velocity dispersions of individual group members and of the lens galaxy closest neighbor (to the East), one could perform a more detail model of the lens potential which would most probably improve the fit.

   We notice that systems with perfect Einstein rings are mostly satisfactorily modeled by a SIE plus the measured external shear (COSMOS~5947+4752 in Fig.~\ref{mmodel}, COSMOS~0124+5121 in Fig.~\ref{mmodel2}, COSMOS~5941+3628 in Fig.~\ref{mmodel3}).  The reason is that, for those lenses, we have arbitrarily chosen the image positions: they are symmetrically distributed around the lens. In addition, the SIE ellipticity is allowed to be null. Hence we are artificially correctly fitting the image positions, whatever the external shear values are (as long as the shear strengths are not too large). For those systems, only the Einstein ring and associated mass are reliable measurements in Table~\ref{SIE}. \\
 For COSMOS~0049+5128, the fit is not good ($\chi^2$=35.3).  If we let  the external shear free, the best fit shear parameters are different than the one measured in  $\S$~\ref{lenscont}.  If we subtract the shear vector given in Table~\ref{knobfig} to the best fit shear vector in Table\ref{SIE} (column 4), we find the direction pointing toward the galaxy closest to the lens (Galaxy 2 in Fig.~\ref{mmodel}). Galaxy 2 is at a projected distance $\sim$13\arcsec\, to the lens galaxy. Would it be at the lens redshift, it would need to have a velocity dispersion $\sigma_v^{Galaxy2}$=174~km~s$^{-1}$ to create the shear necessary to obtain $\chi^{2\prime}$ while fixing the shear parameters to the measured values.  
   
It is the same for the other lenses: if fixed in the lens model, the external shear due to the groups  leads to  $\chi^2>>1$. If we set the shear parameters free,  the best fit shear will point in a direction which correspond to the vectorial summed orientation of the shear  due  to the groups and of the shear due to a secondary (and third in the case of COSMOS~0038+4133) galaxy.  In Table~\ref{shear} we have reported the projected distance between the lens galaxy and the secondary (third) galaxy, as well as the the velocity dispersion  of the second (third) galaxy that is needed to obtain $\chi^{2\prime}$ when fixing the external shear parameters to the one due to the groups. In the case of COSMOS~0050+4901, the velocity dispersion derived for the  main lens and for the secondary galaxy are that of groups rather than that of galaxies:  indeed, when we look at the image of  the lens (Fig.~\ref{mmodel4}), we see that the field is crowded with galaxies. However, the  neighbor density ratio measured in $\S$~\ref{densityratios} do not show any evidence for the presence of a structure at the lens redshift. Therefore if there is actually a group intervening in this system, it should be at a different redshift than that of the main lens galaxy.
For the other systems, the velocity dispersion associated with the secondary lens can be associated with a galaxy mass. For COSMOS~0047+5023, we see  in Fig.~\ref{mmodel3} that the field around the lens is crowded with galaxies: an improved version of the present mass model should take them into account, preferentially using a measurement of their redshift and  velocity dispersion. 

We conclude that  our efforts at measuring the external shear were not in vain as, when using it in the lens model, we clearly identify the missing element : the best fit will point at the closest galaxy to the lens or indicate the realistic presence of a galaxy group in the line-of-sight.  We come back to this result in $\S$~\ref{d4}.

The Einstein radius of the lens galaxy and corresponding mass have been calculated for the best models and are displayed in Table~\ref{SIE}.

\begin{table}
\centering
\caption{\label{shear} Parameters for Galaxy 2.}\begin{tabular}{ c c c  c  }
\hline
\hline
Name   & Distance  & $\sigma_v^{Galaxy2}$& Fig.\\
                 & {\it \arcsec} &{\it km~s$^{-1}$}& \\
\hline
COSMOS~0049+5128&12.8 & 174& \ref{mmodel}\\
COSMOS~5921+0638&1.6& 70 & \ref{mmodel2}\\
COSMOS~0038+4133&3.5 &195 &\ref{mmodel2}\\
             $\prime\prime$                     & 4.6 & 195 &\ref{mmodel2}\\
COSMOS~0047+5023& 3.9 &264 &\ref{mmodel3}   \\
J100140.12+020040.9&1.1&94&\ref{mmodel3} \\
J095930.93+023427.7&4.9&275& \ref{mmodel5}\\
COSMOS~0050+4901&7.4 &366 &\ref{mmodel4} \\
COSMOS~0018+3845&3.5 &113&\ref{mmodel4} \\\
COSMOS~5914+1219&3.0 & 209 & \ref{mmodel4} \\
\hline
\end{tabular}
\tablefoot{ (1) Lens Name. (2) Distance between the lens galaxy and Galaxy 2 (or Galaxy 3 in the case of COSMOS~0038+4133). (3) Velocity dispersion of Galaxy 2 (or Galaxy 3) as explained in $\S$~\ref{siegamma}. (4) Figure number where the lens model is displayed and the Galaxy 2 and 3 are labelled.  }

\end{table}

     \subsubsection{The proportion of DM in the Einstein radius}
     
\begin{figure}[ht]
\begin{center}
\includegraphics[width=8.5cm]{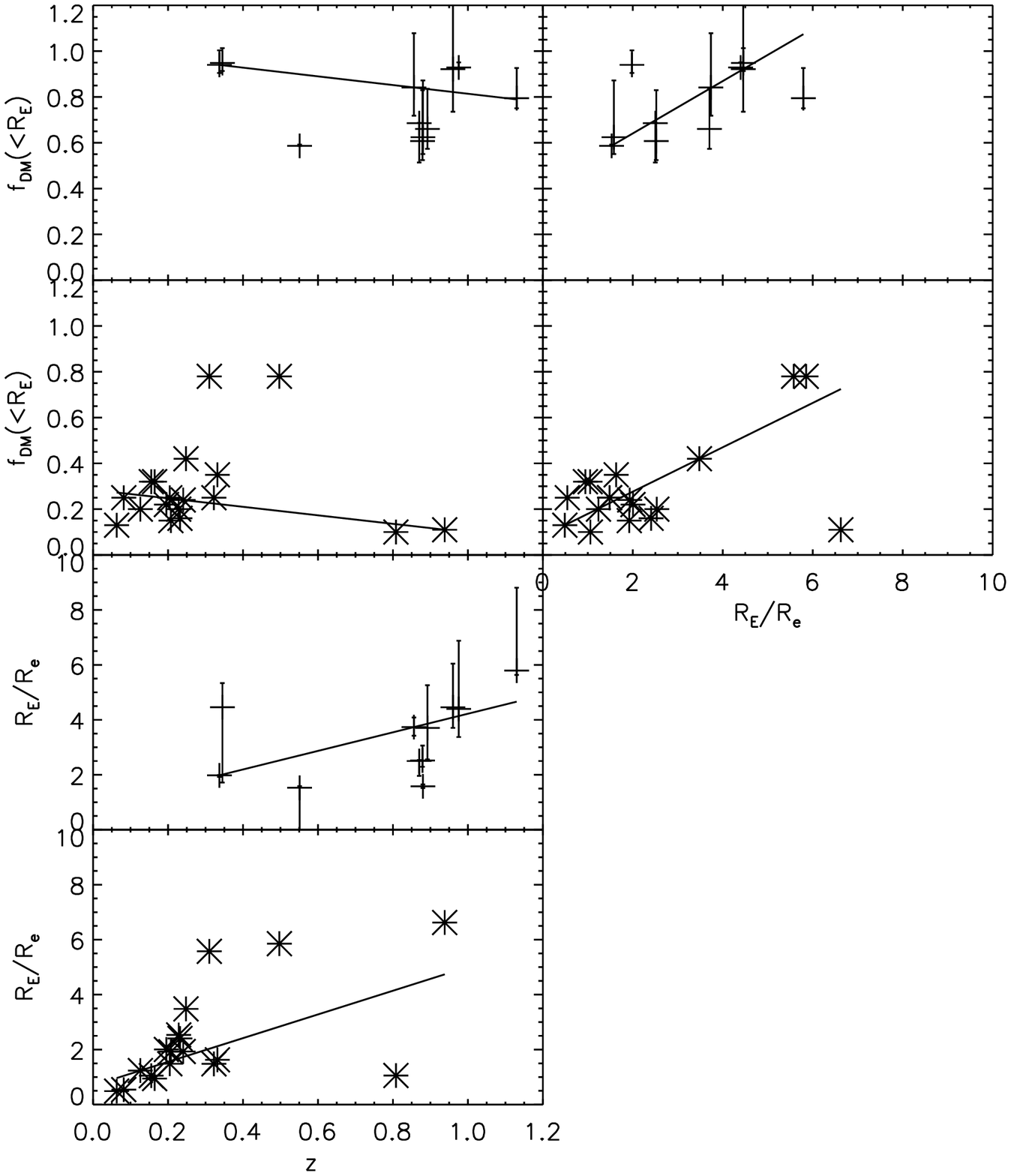}
\caption{Evolution of the dark matter fraction and related parameters for two lens samples: COSMOS (crosses) and JK07 (stars). Top left panels: Evolution of the DM fraction in the Einstein radius with the redshift. The least deviation fits are  displayed (solid lines), and follow the relations  $f_{DM}$($<$R$_E$)=-0.18$\times$z+1.00 for COSMOS and $f_{DM}$($<$R$_E$)=-0.18$\times$z+0.38 for JK07. Top right panels: Variation of the DM fraction as  function of the ratio of the Einstein radius versus effective radius. The least deviation fits give the following relations: $f_{DM}$($<$R$_E$)=+0.11$\times \frac{{\rm R}_E}{{\rm R}_e}$+0.41  (COSMOS) and $f_{DM}$($<$R$_E$)=+0.09$\times \frac{{\rm R}_E}{{\rm R}_e}$+0.08 (JK07).  Bottom panels: Evolution with redshift of the Einstein radius versus effective radius. The  least deviation fits give the following relations: $\frac{{\rm R}_E}{{\rm R}_e}$=+3.4$\times$z+0.8 (COSMOS) and $\frac{{\rm R}_E}{{\rm R}_e}$=+4.3$\times$z+0.7 (JK07). For JK07, we have reported the results obtained for galaxy models without adiabatic compression. The  slopes are similar for models with adiabatic compression.}
\label{fdm}
\end{center}
\end{figure}

 We integrate the galaxy light profile density up to the Einstein radius, and use  the stellar mass of the lens galaxy, identifying the effective radius to the half stellar mass radius, to determine the stellar mass within the Einstein radius, M$_{\star}$($<$R$_E$).  Doing so, we can compare it to the total mass within the Einstein radius,  M($<$R$_E$), obtained during  the lens modeling. This gives us a measurement of  the lens galaxy  projected DM fraction within the Einstein radius: $f_{DM}$($<$R$_E$)= 1-$\frac{{\rm M}_\star{\rm (<R}_E)}{{\rm M(<R}_E)}$. These values are reported in Table~\ref{SIE}. The errors come from the propagation of  the uncertainties in M$_\star$, R$_e$, M($<$R$_E$) and R$_E$. 

The negative DM fraction for COSMOS~0038+4133 may indicate that the stellar mass of the lens galaxy is overestimated. This is surprising as for this system  the photometric and spectroscopic redshifts agree, hence making us confident that the SED, hence the stellar mass, is reliable as well. Another possible explanation is that the light profile fit is not correct and that  we measure too large an effective radius. This is indeed a possible explanation as a bright ring surrounds the lens galaxy and may bias the determination of the light profile. Finally, it is also possible that the source redshift is lower than the one considered here (z$_s^{min}$=1.5). Indeed, would  z$_s$=0.9, the total mass in the Einstein radius would be M($<$R$_E$)=4.88$\times$10$^{11}$M$_\odot$, and the corresponding DM fraction: f$_{DM}$ ($<$R$_E$)=0.6.  Because of all these interrogations, which requires further investigation, we do not keep this system for the rest of the analysis.

 For the 11 remaining systems,  the fraction of DM varies between 0.58$\pm$0.00 
(COSMOS~5921+0638) and 0.95$^{+0.06}_{-0.03}$ (COSMOS~5947+4752).

 In Fig. ~\ref{fdm} we have reported the evolution of f$_{DM}$ ($<$R$_E$) with the redshift. The evolution of  f$_{DM}$($<$R$_E$) with redshift is compatible with constant or slightly decreasing. The interpretation of this tendency is linked to the evolution of the Einstein radius, R$_E$ versus effective radius, R$_e$. Indeed, for a given galaxy  if (1) $R_E \sim R_e$, the fraction of dark matter in the Einstein radius is  expected to be ``low'' as, in this case, we are probing  the region where the galaxy is baryon dominated: f$_{DM}$($<$R$_E$)  is then a lower limit of the total f$_{DM}$. On the contrary, if (2)  $R_E>> R_e$: we are measuring a f$_{DM}$($<$R$_E$) which is getting close to the total  f$_{DM}$, and therefore it is expected to be larger, in average, than the one determined in case (1).  This is indeed the case (see Fig~\ref{fdm}, top right panel): f$_{DM}$($<$R$_E$) increases slowly when $R_E/R_e$ increases.
 From the same Fig.~\ref{fdm} (bottom left panel) we see that the ratio $R_E/R_e$ increases quickly with redshift, meaning that as redshift grows we are measuring f$_{DM}$($<$R$_E$) getting closer to the total f$_{DM}$ of the galaxy.  The fact that  f$_{DM}$($<$R$_E$)  is slightly decreasing or "at best" constant  when the redshift grows suggest that the ``total'' fraction of dark matter is genuinely lower in high redshift lens galaxies, than in the low redshift lens galaxies.\\
 
 Interestingly, we observe the same tendency if we consider the lens sample from Jiang and Kochanek (2007, JK07 hereafter): they measure the DM fraction (assuming or not adiabatic compression in their galaxy models and Salpeter initial mass function (IMF))  in the Einstein radius of 22 galaxies spanning a lens redshift $z_l$=[0.0808,1.004].  If we look at the 18 galaxies with  $R_E/R_e <$10 as in our sample (see Fig.~\ref{fdm}), we notice that, similarly to our sample: (i) f$_{DM}$($<$R$_E$) decreases slightly  with redshift with the same slope than for our sample (-0.18), (ii) the radius $R_E/R_e$ increases with redshift and (iii) f$_{DM}$($<$R$_E$) slightly increases when $R_E/R_e$ increases. \\
 We add that, even if the two samples show similar behavior, we chose not to mix them for the display and slope calculations in order to avoid misinterpretations due to possible systematics affecting the measurement of the DM fraction in the two different methods  (e.g. different IMFs).

\section{Discussion and conclusions}\label{disc}

From the analysis of the COSMOS strong lenses in the redshift range
0.34 to 1.13, we get three major results: (1) the lens galaxy stellar masses increases with the redshift, (2) the lens galaxy environments are compatible with those of ETGs in this redshift range and (3) the DM fraction of the lens galaxies in the Einstein radius slightly decreases with the redshift, when the ratio between the Einstein radius versus the effective radius strongly increases with redshift. Let us discuss each one of these trends.

\subsection{Increase of the lens stellar mass with redshift}\label{d1}

 In $\S$~\ref{smlg}, we have  measured that the lens galaxy stellar mass increases with redshift.
Is this a  result of the selection method used to built the lens sample?  At z$\sim$0.4, one of our lenses has an effective radius of $\sim$0.3\arcsec\, and at z$\sim$0.9, the smallest lens galaxy effective radius is $\sim$0.2\arcsec (see Table~\ref{priors}). The effective radius is a lower limit for the  Einstein radius. These ``lower limit Einstein radii'' correspond to lower limits in mass within the Einstein radius that are in our case: 10$^{10.9}$ M$_\odot$ at z=0.4 and 10$^{11.0}$ M$_\odot$  at z=0.9, assuming a source at z=1.5.
  Hence, for a given source redshift, the minimal lens galaxy mass  should increase with the lens redshift in order to be visually detected;  hence, most probably, the minimal stellar mass might increase as well with redshift. Following this reasoning,  we expect the sample average stellar mass  to increase with the redshift. While this reasoning is simplistic as the source redshifts are different for every lens,  it still explains partially the increasing stellar mass with increasing redshift of the lens galaxies.
So far, it is not clear whether this "detection bias" effect is the only reason for the lens galaxy stellar mass increase with redshift. 
This result should be investigated in dedicated numerical simulations (e.g. van de Ven et al. (2009), Mandelbaum et al. (2009)).

\subsection{Evolution of the lens galaxy environment}\label{d2} 

 Our study of the environments of COSMOS lens galaxies,
via the projected galaxy number density in $\S$~\ref{voisin},
indicates that the environment of  lens galaxies is similar to that of ETGs across the whole range of redshifts tested here. This result extends the previous measurements made in the redshift range [0.068; 0.513] with the SLACS sample (T09 and Auger 2008). It is particularly interesting as the COSMOS and SLACS samples have very different selection criteria: hence this result is most probably  a genuine characteristic of strong lens galaxies rather than a selection bias.

  We also notice that both neighbor density estimator tested here appear to be reliable estimators of the environment of  galaxies, 
 as X-ray groups and clusters are actually detected  around lenses with high $\frac{\Sigma_{10}}{<\Sigma_{10}>_t}$ and high $\frac{D_{1}}{<D_{1}>_t}$ ratios.    
\subsection{The decreasing dark matter fraction with redshift}\label{d3}

  We have measured the lens galaxy DM fraction in the Einstein radius by combining the  total mass in the Einstein radius, the light density profile and the stellar mass in the galaxy.  The projected DM fraction decreases with redshift  even though the ratio between the Einstein radius and the effective radius increases with the redshift (see Fig.~\ref{fdm}).  A similar trend is seen in the lens galaxy dataset of JK07 which covers a similar redshift range. 
Using toy models and a $\lambda$CDM cosmology and the SLACS lenses, Napolitano et al. (2010)  and Tortora et al. (2010) first notice that,  for a fixed stellar mass and age, the DM fraction is similar for high and low redshift galaxies. They also notice that the slope of the distribution of  f$_{DM}$ versus age  is steeper than explained by their model and they invoke different scenarios to interpret the discrepancy (including adiabatic compression  and different IMF for different galaxy ages). Another possible effect responsible for the DM fraction decrease with redshift is the increasing stellar density of ETGs with redshift as discussed by Bezanson et al. (2009). But to understand if the results are affected by one of this effect, f$_{DM}$ need to be measured in comparable radius, not different for each galaxy, contrary to  the Einstein radius. 
In order to conclude on the favorite ongoing processes on ETGs since z$\sim$1  as a function of their age and stellar mass, a joint analysis of the three lens galaxy samples would certainly be a first step towards a better understanding.
 
  \subsection{Some words on the lens models}\label{d4}
   Using lens modeling  we measure the  Einstein radii and related total masses for the lens galaxies. Doing so, we were able to measure their DM fractions within their Einstein radius. In addition, the lens models were used to test the calculation of the external shear due to the groups around the lens galaxies. We observe that, in every lens model,  the difference between the best fit external shear and the shear due to the groups  points towards  the closest  galaxy to the lens. We calculate that, those secondary galaxies need  to have realistic velocity dispersions for the lens models to provide good fits.
    
Moreover,  we have measured that a catalog of  high mass groups and clusters  modeled by TIS  is giving very similar total shear strengths and orientations than  a combination of  high mass and low mass group catalogs.
    Therefore we conclude that a measurement of the global external shear affecting the lens potential at the lens galaxy location could be made if one could combine : 1) a measurement of the  redshift and velocity dispersion of the lens galaxy closest neighbor(s) plus 2) the locations,  masses and radii of the most massive groups and clusters (such as the one provided by the X-ray observations in the COSMOS field) in a $\gtrsim$5\arcmin\, radius around the lens galaxy. By fixing the external shear  one would break an important source of degeneracy in the lens models. 
   
\subsection {Conclusions}
 On one hand, we have measured that the environment of lens galaxies is similar to that of non ETGs over a wide redshift range: between 0.068 and 1.13 if we put together results from the SLACS sample (T09, Auger 2008) and from the  COSMOS sample (this paper).
 On the other hand, we have built up an ensemble of clues suggesting that the mass properties of
lens galaxies evolve
with redshift.

Indeed, we notice that at high redshift, lens galaxies have
a large stellar mass  and a  total dark matter fraction $<f_{DM}>\sim$0.7 (for z$>$0.8). On the
contrary, at low redshifts, lens galaxies have lower stellar mass and their DM fraction are $<f_{DM}>$~$>$0.8.

This results advocates in favor of high stellar density of high redshift ETGs in comparison to low redshift ETGs as suggested by the results of Bezanson et al. (2009). Or it could be that the difference between low and high redshift lens galaxy population is a consequence of the stellar population aging  and different IMFs at different ages (see Napolitano et al. 2010). 
 It could also be that the effects measured here are related to the lensing efficiency, which in this case would be a complex combination
of (1) the lens population number density, (2) the source population
distribution in space, redshift and luminosity, (3) the survey
properties (sensitivity, band, size, angular resolution), (4) biases in the lens sample selection.

To disentangle between an evolutionary or a pure lensing origin of the effects discovered in this study, the evolution of lens galaxy properties with redshift needs to be
studied in dedicated numerical simulations (e.g. van de Ven et
al. (2009), Mandelbaum et al. (2009)). Whether the effects are 
intrinsic to the massive early type galaxy population or to the lensing efficiency, they must be fully understood if one
wants to study properly the galaxy properties gathered from lens
galaxy populations.

\begin{acknowledgements} We acknowledge the anonymous referee for providing a detailed and very useful  report. We are gratefully indebted  to M. Limousin and R. Gavazzi for
  enlightening discussions. DA thanks CNRS and CEA for support and
  visits to the Geneva observatory where this work was finalized.  JPK
  thanks support from CNRS and SL2S ANR-06-BLAN-0067 and DESIR ANR-07-BLAN-0228. E.J. acknowledges the support of the NPP, administered by Oak Ridge
Associated Universities through a contract with NASA.
\end{acknowledgements}

\end{document}